\documentclass{aa}  

\usepackage{graphicx}
\usepackage{txfonts}

\usepackage{natbib}
\bibpunct{(}{)}{;}{a}{}{,} 
\usepackage[colorlinks=true,linkcolor=blue,citecolor=blue]{hyperref}
\usepackage{amsmath}

\usepackage{tkz-graph}  
\usetikzlibrary{shapes.geometric, positioning}%

\usepackage{subfig}
\usepackage[shortlabels]{enumitem}

\newcommand{\fesc}{f_\mathrm{esc}}
\newcommand{\Robs}{\mathcal{R}_\textrm{obs}}
\newcommand{\Rint}{\mathcal{R}_\textrm{int}}
\newcommand{\trans}{e^{-\tau^\textrm{HI}_\mathrm{LyC}}}
\newcommand{\optdepth}{\tau^\textrm{HI}_\mathrm{LyC}}
\newcommand{\Auv}{A_\mathrm{UV}}

\newcommand{\nuis}{\boldsymbol{\theta}}
\newcommand{\pop}{\boldsymbol{\phi}}

\newcommand{\fesci}{f_{\mathrm{esc},i}}
\newcommand{\Robsi}{\mathcal{R}_{\textrm{obs},i}}
\newcommand{\Rinti}{\mathcal{R}_{\textrm{int},i}}
\newcommand{\transi}{e^{-\tau^\textrm{HI}_U}_i}
\newcommand{\Auvi}{A_{\mathrm{UV},i}}

\newcommand{\nuisi}{\boldsymbol{\theta}_i}

\newcommand{\Nexp}{10,000 }
\newcommand{\BegleyBayesfesc}{0.05 \pm 0.02}

\newcommand{\ConstMuVal}{0.05^{+0.01}_{-0.01}}
\newcommand{\ConstAll}{\pop_\mathrm{const}=\mu=\ConstMuVal}
\newcommand{\ExpMuVal}{0.05^{+0.01}_{-0.02}}
\newcommand{\ExpAll}{\pop_\mathrm{exp}=\mu=\ExpMuVal}
\newcommand{\LognormMuVal}{0.00^{+0.33}_{-0.00}}
\newcommand{\LognormSigmaVal}{2.63^{+0.06}_{-0.12}}
\newcommand{\LognormAll}{\pop_\mathrm{lognorm}=(\mu=\LognormMuVal, \sigma=\LognormSigmaVal)}
\newcommand{\BimodalFracVal}{0.84^{+0.10}_{-0.05}}
\newcommand{\BimodalMuVal}{0.29^{+0.24}_{-0.21}}
\newcommand{\BimodalSigmaVal}{0.11^{+0.05}_{-0.11}}
\newcommand{\BimodalAll}{\pop_\mathrm{bimodal}=(w=\BimodalFracVal, \mu=\BimodalMuVal, \sigma=\BimodalSigmaVal)}
\newcommand{\BlueExpMuVal}{0.14^{+0.04}_{-0.06}}
\newcommand{\RedExpMuVal}{0.01^{+0.01}_{-0.00}}

\begin{document} 

   \title{Inferring the Distribution of the Ionising Photon Escape Fraction}
   \subtitle{}

   \author{Kimi C. Kreilgaard
          \inst{1}
          \and
          Charlotte A. Mason\inst{1}
          \and
          Fergus Cullen\inst{2}
          \and 
          Ryan Begley\inst{2}
          \and
          Ross J. McLure\inst{2}}

   \institute{The Cosmic Dawn Center, 
                Niels Bohr Institute, 
                University of Copenhagen,
              Jagtvej 128, DK-2200 Copenhagen N, Denmark\\
         \and
             Institute for Astronomy, 
             University of Edinburgh,
             Royal Observatory, Edinburgh EH9 3HJ\\
             }

   \date{Received MONTH DATE, YEAR; accepted MONTH DATE, YEAR}

 
  \abstract{
The escape fraction of ionising photons from galaxies ($f_\mathrm{esc}$) is a key parameter for understanding how intergalactic hydrogen became reionised, but it remains mostly unconstrained. Measurements have been limited to the average value in galaxy ensembles and handfuls of individual detections. To help understand which mechanisms govern ionising photon escape, here we infer the distribution of $f_\mathrm{esc}$. We develop a hierarchical Bayesian inference technique to estimate the population distribution of $f_\mathrm{esc}$ from the ratio of Lyman Continuum to non-ionising UV flux measured from broadband photometry. We apply it to a sample of 148 $z \simeq 3.5$ star-forming galaxies from the VANDELS spectroscopic survey. We explore four physically motivated distributions: constant, log-normal, exponential and bimodal, recovering $\langle f_\mathrm{esc} \rangle \approx5\%$ for most models. 
We find the observations are best described by an exponential $f_\mathrm{esc}$ distribution with scale factor $\mu=0.05^{+0.01}_{-0.02}$.
This indicates most galaxies in our sample exhibit very low escape fractions while predicting substantial ionising photon leakage for only a few galaxies, implying a range of optical depths in the ISM and/or time variability in ionising photon escape. 
We rule out a bimodal distribution at high significance, indicating that a purely bimodal model of ionising photon escape (due to very strong sightline and/or time variability) is not favoured. We compare our recovered exponential distribution with the SPHINX simulations and find that, while the simulation also predicts an exponential-like distribution, it significantly underpredicts our inferred mean. 
The distribution of $f_\mathrm{esc}$ can be a vital test for simulations in understanding ionising photon leakage and is important to consider to gain a complete picture of reionisation.
}
   \maketitle



\section{Introduction}
\label{sec:intro}

The Epoch of Reionisation (EoR) refers to the transformation of the predominantly neutral hydrogen present after recombination to the ionised intergalactic medium (IGM) that we observe today. Constraining the EoR is therefore fundamental for understanding the large-scale evolution of the IGM and the sources of reionisation. Observations have estimated that the reionisation of the IGM is completed within a billion years after the Big Bang, ending at $z\sim 5.5$ and on-going at $z\sim8$ \citep[e.g.,][]{fanObservationalConstraintsCosmic2006, schroederEvidenceGunnPeterson2013, mcgreerModelindependentEvidenceFavour2015, masonUniverseReionizingBayesian2018, bosmanConstraintsMeanFree2021, qinReionizationGalaxyInference2021, bolanInferringIntergalacticMedium2022}.

Ionising intergalactic hydrogen requires a vast injection of high-energy photons into the IGM. The primary sources of these photons are still hotly debated. While active galactic nuclei (AGN) are strong emitters of hydrogen ionising photons at lower redshifts, the number density of UV-bright AGN drops rapidly with increasing redshift \citep{airdEvolutionXrayLuminosity2015, parsaNoEvidenceSignificant2018, mcgreerVizieROnlineData2018, kulkarniEvolutionAGNUV2019, faisstJointSurveyProcessing2022}, and they are therefore not expected to play a major role in hydrogen reionisation \citep[e.g.,][]{mattheeLittleRedDots2024,dayalUNCOVERingContributionBlack2024}. Thus, star-forming galaxies are the most likely sources of the bulk of the ionising photons \citep[e.g.,][]{robertsonCOSMICREIONIZATIONEARLY2015,bouwensREIONIZATIONPLANCKDERIVED2015}. The James Webb Space Telescope (JWST) is confirming that a majority of $z>6$ galaxies appear to be dominated by the light from young, massive stars which can produce hydrogen ionising photons \citep[e.g.,][]{endsleyJWSTNIRCamStudy2023,endsleyStarformingIonizingProperties2023,prieto-lyonProductionIonizingPhotons2023,rinaldiMIDISUnveilingRole2023}.
However, which physical properties of galaxies are most conducive to the production and escape of ionising photons remains uncertain. The relative importance of stellar and/or dark matter halo mass, UV magnitude, specific star formation rate (SSFR), star formation rate (SFR) `burstiness', dust, and stellar populations and radiation fields, among others, are still debated in both observational and theoretical work \citep[e.g.,][]{paardekooperFirstBillionYears2015,finkelsteinConditionsReionizingUniverse2019,maNoMissingPhotons2020,naiduRapidReionizationOligarchs2020,mattheeReSolvingReionization2022,fluryLowredshiftLymanContinuum2022}.

Measuring the ionising photon output from early galaxies requires constraining three components: (i) the number of galaxies producing ionising photons, (ii) the efficiency with which they produce ionising photons and (iii) the fraction of the produced ionising photons that escape from the galaxies into the IGM \citep[e.g.][]{gnedinModelingCosmicReionization2022, robertsonGalaxyFormationReionization2022}. This paper addresses the latter quantity, by proposing a novel technique to infer the \emph{distribution} of the absolute, hydrogen ionising photon escape fraction, $\fesc$, of early galaxies. The escape fraction we will refer to (unless otherwise stated) is sight-line dependent and thus distinguishes itself from angle-averaged values typically reported for simulations (see \citealp{simmondsImpactNebularLymanContinuum2024} for more details on the nuances of escape fraction definitions).

Independent of the adopted definition, the escape fraction remains challenging to constrain due to the difficulty of directly observing Lyman continuum (LyC) flux from distant galaxies. A measurement involves detecting the flux blue-ward of 912\AA, but this flux is typically very low due to low escape fractions and absorption by residual neutral hydrogen in the IGM attenuating the emitted LyC flux \citep{robertsonEarlyStarformingGalaxies2010}. Measurements are further complicated by the fact that the IGM is optically thick to ionising photons above redshift $z \sim 4$ \citep{madauRadiativeTransferClumpy1995, inoueUpdatedAnalyticModel2014} and therefore have to be carried out at lower redshifts. This also implies that direct estimates of the escape fraction are not possible within the EoR.

Measurements of the LyC escape fraction have been made for a few dozen individual galaxies in, respectively, the local Universe \citep[e.g.,][]{izotovLowredshiftLymanContinuum2018, izotovLymanContinuumLeakage2021, fluryLowredshiftLymanContinuum2022a} and at intermediate redshift ($2>z<4.5$) \citep[e.g.,][]{vanzellaDETECTIONIONIZINGRADIATION2012, mostardiHIGHRESOLUTIONHUBBLESPACE2015,vanzellaHUBBLEIMAGINGIONIZING2016, debarrosExtremeIIIEmitter2016, shapleyQ1549C25CLEANSOURCE2016, bianHighLymanContinuum2017, vanzellaDirectLymanContinuum2018, fletcherLymanContinuumEscape2019, rivera-thorsenGravitationalLensingReveals2019, marques-chavesExtremeBlueNugget2022, saxenaNoStrongDependence2022, keruttLymanContinuumLeaker2023}, a majority of which exhibit high escape fractions ($\fesc>0.2$), and are, because of this property, known as LyC leakers. Another approach frequently employed is to estimate the average escape fraction from larger samples of galaxies, a method which incorporates non-LyC leakers to investigate population-averaged values. Such studies suggest the average escape fraction does not exceed a few per cent \citep[e.g.,][]{steidelKeckLymanContinuum2018, pahlUncontaminatedMeasurementEscaping2021, mestricUpperLimitsEscape2021, begleyVANDELSSurveyMeasurement2022}.

Individual detections and/or average escape fraction estimates are, however, insufficient for determining which physical processes govern the escape of ionising photons. Observations of the escape fraction are limited to the specific line of sight with which the galaxies are observed (except for rare cases where a lensed object grants observations of multiple sight lines e.g. \citealp{rivera-thorsenGravitationalLensingReveals2019}). Furthermore, when studying distant galaxies, we are left with poor or no spatial resolution. Both are limiting factors for our understanding of how ionising photons escape their host galaxy using individual detections, while an estimate of the average escape fraction only informs of the amount of photons escaping and similarly does not reveal how they escape.

Lyman-continuum leakage has been shown to correlate with tracers of concentrated star formation and young, highly ionising, stellar populations, such as star formation rate (SFR) surface density and O32 ratio, and tracers of low line-of-sight gas and dust column density, e.g. blue UV $\beta$-slopes and strong Ly-$\alpha$ emission \citep[][]{jaskotORIGINOPTICALDEPTH2013,izotovDetectionHighLyman2016,fluryLowredshiftLymanContinuum2022,chisholmFarultravioletContinuumSlope2022}. These observational trends imply hard radiation fields and stellar feedback may be crucial for LyC escape.
In the commonly held picture, there are two main mechanisms capable of producing LyC leakage from star-forming regions: either LyC photons escape through (i) an ionisation-bounded nebula with "holes" or (ii) a density-bounded nebula \citep[e.g.,][]{zackrissonSPECTRALEVOLUTIONFIRST2013, reddyCONNECTIONREDDENINGGAS2016, steidelKeckLymanContinuum2018, kimmUnderstandingEscapeLyC2019, kakiichiRadiationHydrodynamicsTurbulent2021}.
In the first scenario, the leakage of LyC photons is caused by stellar winds or supernova feedback opening up low-density channels in the neutral interstellar medium (ISM). These channels allow for the effective escape of LyC photons, which would otherwise be absorbed by the surrounding neutral region - leading to a strong sight-line dependence on the escape fraction. In the density-bounded picture, the LyC flux from a very powerful star-formation episode “exhausts” all the neutral hydrogen before a complete Strömgren sphere can form, thereby allowing LyC photons to escape into the IGM, exhibiting a more symmetric escape of the ionising photons from the star-forming region. Galaxies are naturally more complex than these two simple scenarios, in particular due to time variability of star formation, and LyC leakage is therefore likely the result of mixtures of the two mechanisms. 

One way to explore the prevalence of the two different leakage mechanisms (and variations thereof) is through simulations of reionisation \cite[see e.g.,][]{kimmESCAPEFRACTIONIONIZING2014, paardekooperFirstBillionYears2015, barrowLymanContinuumEscape2020, maNoMissingPhotons2020, rosdahlLyCEscapeSPHINX2022, choustikovPhysicsIndirectEstimators2023, kostyukIonizingPhotonProduction2023}. Large cosmological simulations have the advantage of a fully accessible 3-dimensional view of a galaxy and thus multiple sight lines available for each galaxy. Furthermore, the intrinsic spectrum and various galaxy properties are known such that the escape fraction can be calculated directly at all times along with any correlations between the escape fraction and physical properties. These simulations also predict that the escape fraction fluctuates strongly in individual galaxies over time-scales of a few Myr.

Which physical mechanisms dominate the escape of ionising photons, how sightline-dependent the leakage is, and how much it fluctuates with time are all things that will affect the distribution of the ionising photon escape fraction. Measuring the distribution of escape fractions across a population of galaxies and comparing it with predictions from simulations, will therefore provide new insight into the nature of how ionising photons escape \citep[e.g.,][]{cenQUANTIFYINGDISTRIBUTIONSLYMAN2015}, and is the aim of this study.

We develop a method for estimating the overall distribution of escape fractions for a larger galaxy population. We build on previous work by \cite{begleyVANDELSSurveyMeasurement2022}, who inferred the most likely escape fraction to describe a large sample of galaxies to be $\langle \fesc \rangle=0.05\pm0.01$ using a Bayesian inference framework and by \citet{vanzellaGREATOBSERVATORIESORIGINS2010} who considered different models for the escape fraction distribution. We extend the framework devised by \citet{begleyVANDELSSurveyMeasurement2022} to that of a hierarchical Bayesian inference scheme, which will enable us to infer a population distribution, and we perform a model comparison to assess which distribution shape best fits the observations. 

This paper is structured as follows. Section \ref{sec:data} describes the datasets used in this study, focusing on the sample selection and the measurement of the LyC to non-ionising UV flux ratios. In Section \ref{sec:modelling}, we introduce a model linking the escape fraction to the observable LyC to non-ionising UV flux ratio. Section \ref{sec:bayes_inference} details the hierarchical Bayesian inference framework that we adopt to infer the population distribution of escape fractions across our galaxy sample. We present our constraints on the escape fraction distribution in Section \ref{sec:results}, before discussing the significance of our results in Section \ref{sec:discussion} and summarising our conclusions in Section \ref{sec:conclusion}.

Throughout this paper, we adopt cosmological parameters $H_0 = 70 \mathrm{km s}^{-1}\mathrm{ Mpc}^{-1}$, $\Omega_\mathrm{m}=0.3$ and $\Omega_\Lambda=0.7$. Unless otherwise stated, we refer to the absolute, line-of-sight, escape fraction of LyC photons, denoted by $\fesc$, as simply the escape fraction.
\section{Data and Sample Selection}
\label{sec:data}

This study utilised observations of star-forming galaxies within the Chandra Deep Field South (CDFS) extracted from three key datasets: (i) a sample of spectroscopically confirmed galaxies provided by the final data release (DR4) of the VANDELS ESO public spectroscopy survey \citep{mclureVANDELSESOPublic2018,pentericciVANDELSESOPublic2018,garilliVANDELSESOPublic2021}, (ii) ultra-deep VIMOS U-band imaging covering the rest-frame LyC \citep{noninoDEEPBANDIMAGING2009} and (iii) imaging of the CDFS in the HST ACS F606W (hereafter V606) filter, released as part of version 2.0 of the Hubble Legacy Field programme \citep{whitakerHubbleLegacyField2019}, covering the non-ionising UV flux.

The final sample selection is outlined in Section \ref{sub:sample_selection}, while Section \ref{sub:robs} presents our fundamental observable: the observed LyC to non-ionising UV flux ratio, $\Robs$. The galaxy sample used in this study was originally presented in \cite{begleyVANDELSSurveyMeasurement2022}, where more details are available.

\subsection{Sample Selection}
\label{sub:sample_selection}

The VANDELS ESO spectroscopy survey obtained ultra-deep (20-80 hours of integration), red optical ($4800 \mathrm{Å}<\lambda_\mathrm{obs}<10200\mathrm{Å}$) spectra for a sample of 2087 galaxies in the redshift range of $1.0\leq z_\mathrm{phot}\leq 6.4$ \citep{mclureVANDELSESOPublic2018,pentericciVANDELSESOPublic2018,garilliVANDELSESOPublic2021}.

\begin{figure}
    \centering
    \includegraphics[width=\textwidth/2]{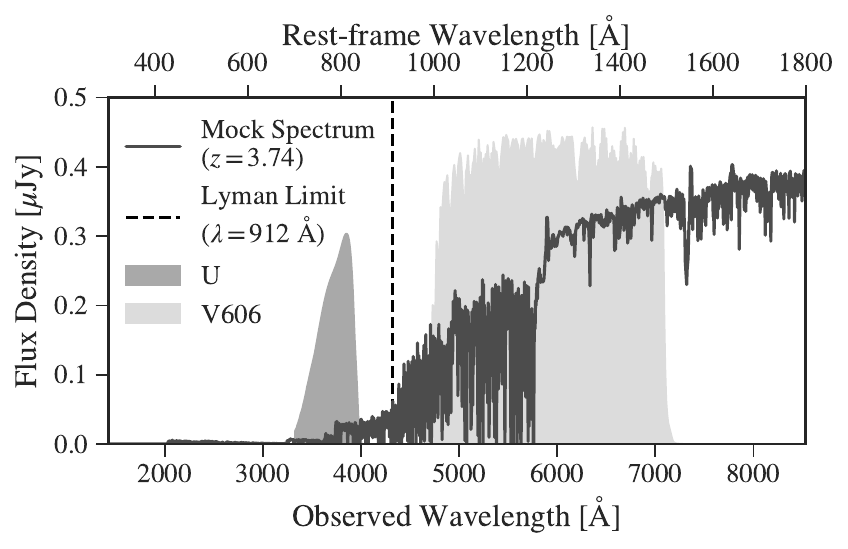}
    \caption{Illustration of a mock spectrum with the U-band (LyC) and V606 (non-ionising UV) filters marked by the shaded areas. The Lyman limit is indicated with the vertical, dashed, black line. The fundamental observable, the LyC to non-ionising UV flux ratio, $\Robs$, (see Section \ref{sub:robs}) measures the flux captured by the U-band (left) divided by the flux captured by the V606 filter (right). To ensure the U-filter is always capturing flux short-ward of the Lyman limit, and thus measuring ionising photons, we impose a lower redshift bound of $z=3.35$ on the galaxy sample. This mock spectrum is obtained from the intrinsic BPASS spectrum that was fitted to the stacked VANDELS spectrum in \cite{begleyVANDELSSurveyMeasurement2022} (see Section \ref{sub:rint}). The flux below the Lyman limit has been attenuated according to an assumed $\fesc=20 \%$, whilst dust attenuation has been applied to the non-ionising region and the full spectrum has been multiplied with a simulated IGM transmission (exhibiting a transmitted fraction of 14\% through the U filter) (see Section \ref{sub:trans}).}
    \label{fig:toy_spectrum}
\end{figure}

For this project, we are interested in star-forming galaxies, with high-quality (reliable at the 99 per cent level) spectroscopic redshift within the interval $3.35\leq z_\mathrm{spec} \leq 3.95$. Since the IGM is optically thick to ionising photons above redshift $z \sim 4$ \citep{madauRadiativeTransferClumpy1995, inoueUpdatedAnalyticModel2014}, the upper limit represents the highest redshift, and thus the time closest to the EoR, that we can observe galaxies that still allow the direct detection of LyC emission. We require a robust measurement of the redshift to properly represent the line of sight transmission through the IGM, as well as be able to accurately measure the flux in the ionising region of the spectrum. Furthermore, we impose a lower bound on the redshift of galaxies of interest, to ensure that the U-band filter only samples rest-frame wavelengths short-ward of the Lyman limit, i.e. photons with enough energy to ionise hydrogen. Together, these requirements restrict the initial galaxy sample to 242 galaxies. In Figure \ref{fig:toy_spectrum}, we show the response curves for the U and V606 filters together with the spectrum of a mock galaxy at redshift $z=3.74$.

Lastly, targets with significant contamination in the imaging data from nearby companion objects were removed and potential AGNs were excluded, reducing the final sample to 148 galaxies. The distribution of redshifts is mostly uniform within the selected range displaying a median of $z=3.58$, while the distribution of stellar mass exhibits a median of $\log(M_*/M_\odot)=9.5$ with a range spanning from 8.6 to 10.5. For more details regarding the final sample selection and cleaning, we refer to \cite{begleyVANDELSSurveyMeasurement2022}, where the distributions of redshift and stellar mass for the final sample are also displayed.

\begin{figure}[]
    \centering
    \includegraphics[width=\textwidth/2]{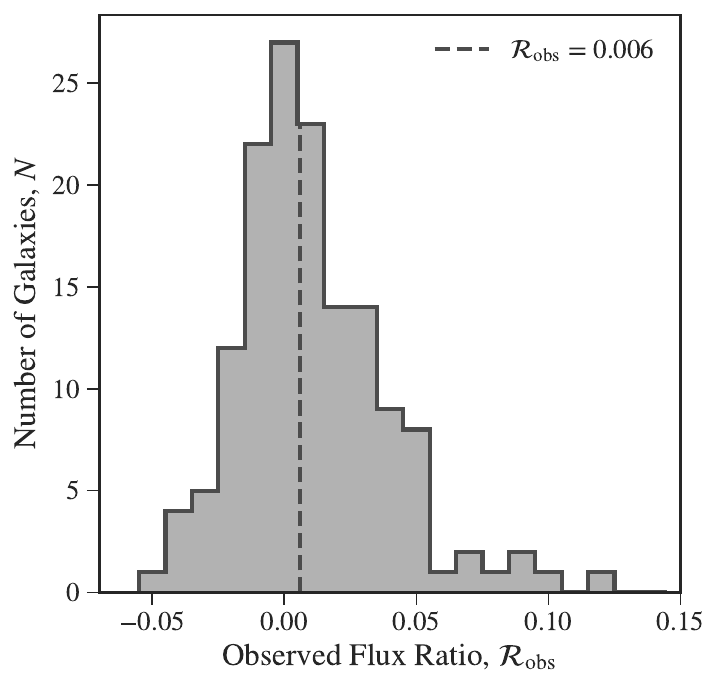}
    \caption{The distribution of the observed LyC to non-ionising UV flux ratio, $\Robs$, exhibited by our final sample of 148 star-forming galaxies (see Section \ref{sec:data}). The dashed line marks the median observed flux ratio of the sample. This distribution constitutes the observed data in our Bayesian inference framework (See Section \ref{sec:bayes_inference}), and thus provides the fundamental observational constraint on estimates relating to the escape fraction. Figure adapted from \cite{begleyVANDELSSurveyMeasurement2022}.}
    \label{fig:robs_distribution}
\end{figure}

\subsection{The Observed LyC to Non-Ionising UV Flux Ratio}
\label{sub:robs}

The observed LyC to non-ionising UV flux ratio, $\Robs$, is obtained with imaging in the U-band and the V606 filter, and defined as:
\begin{equation}
    \Robs = \left( \frac{L_\mathrm{LyC}}{L_\mathrm{UV}} \right)_\mathrm{obs} = \left( \frac{\langle f_\mathrm{U} \rangle }{\langle f_\mathrm{V606} \rangle} \right)_\mathrm{obs},
    \label{eq:robs}
\end{equation}
where $\langle f_\mathrm{U} \rangle$ and $\langle f_\mathrm{V606} \rangle$ are the flux densities per unit frequency measured within 1.2-arcsec diameter apertures. In Figure \ref{fig:toy_spectrum}, this corresponds to the flux captured in the dark grey filter (below the Lyman limit) divided by the flux captured in the light grey filter (above the Lyman limit).

We refer to \cite{begleyVANDELSSurveyMeasurement2022} for details regarding the preparation of the imaging data, which includes PSF-homogenisation, astrometry calibration and sky-subtraction. These steps are taken to ensure that the aperture photometry can be directly compared between the two filters, which entails the apertures capturing the same fraction of total flux for each object.

The distribution of $\Robs$ for the final sample of 148 star-forming galaxies is displayed in Figure \ref{fig:robs_distribution}. Within our framework, it is this distribution that provides the main observational constraint on estimations of the escape fraction and, consequently, the distribution of the escape fraction across the galaxy population.
\section{Modelling Framework}
\label{sec:modelling} 
In Section \ref{sub:gen_model}, we introduce a forward model which relates the escape fraction to the observable LyC to non-ionising UV flux ratio, $\Robs$ (see Section \ref{sub:robs}). This model requires a set of additional parameters, $\nuis=(\Rint, \trans, \Auv)$, which are related to, respectively, the intrinsic ionising photon emission, the optical depth for ionising photons along a given sight line through the IGM, and dust attenuation in the non-ionising UV. These parameters are described separately in sections \ref{sub:rint}, \ref{sub:auv} and \ref{sub:trans}.

\subsection{Forward Model for the Escape Fraction}
\label{sub:gen_model}

We employ the physically motivated, forward model presented by \cite{begleyVANDELSSurveyMeasurement2022}, which relates the observable flux ratio, $\Robs$, to the escape fraction of ionising photons, $\fesc$:
\begin{equation}
    \Robs = \fesc \cdot \Rint \cdot \trans \cdot 10^{0.4 \Auv},
    \label{eq:gen_model}
\end{equation}
where $\Robs$ and $\Rint$ are, respectively, the observed and intrinsic LyC to non-ionising UV flux ratios (see Equation \ref{eq:robs}), $\trans$ is the line-of-sight, transmitted fraction of ionising photons through the IGM and CGM integrated through the U-band filter, and $\Auv$ is the dust attenuation measured at the effective wavelength of the V606 filter.

In Figure \ref{fig:toy_spectrum}, we have provided a mock spectrum which helps visualise how the different components of Equation \ref{eq:gen_model} affect $\Robs$ compared to its intrinsic counterpart, $\Rint$. The escape fraction will reduce the observed flux at wavelengths below the Lyman limit ($\simeq912$Å) (decreasing $\Robs$). Similarly, the absorption of ionising photons by the IGM and CGM, accounted for in the model by an integrated transmitted fraction, $\trans$, will affect only flux measured in the ionising region (decreasing $\Robs$). In contrast, the UV dust attenuation will reduce the observed flux at wavelengths above the Lyman limit (increasing $\Robs$). The forward model intrinsically assumes that dust attenuation of ionising photons is contained in the escape fraction parameter, $\fesc$, and thus $\Auv$ only impacts the non-ionising flux.

Modelling the escape fraction through a flux \emph{ratio} between the LyC and the non-ionising UV band reduces the modelling complexity, eliminating the need to find the normalisation of the star-formation history (SFH) when fitting an intrinsic spectral energy distribution (SED). Furthermore, the scaling ensures that the inference of the escape fraction is comparable for galaxies at (slightly) different redshifts since the resulting displacement of the rest-frame wavelength of the U-band and V606 filter, and thus the energies of the photons captured, will affect both filters similarly. Anchoring the observation of the LyC flux, which is prone to absorption from intervening neutral hydrogen, to another more readily detectable region of the spectrum thus makes for a more reliable measurement. The ionising photons in the LyC are predominantly produced by OB stars, whose spectra are known also to dominate the UV region. Fluxes from these regions are therefore closely related, and a ratio between the two becomes practical for the analysis.

\subsection{The Intrinsic LyC to Non-Ionising UV Flux Ratio}
\label{sub:rint}

The intrinsic LyC to non-ionising UV flux ratio, $\Rint$, is defined similarly to its observed counterpart (see Section \ref{sub:robs} and Equation \ref{eq:robs}). It is computed from the intrinsic SED, whose shape is determined by the properties of the underlying stellar population.
In the analysis of \citet{begleyVANDELSSurveyMeasurement2022}, $\Rint$ was determined by fitting Binary Population and Spectra Synthesis (BPASS) \citep{eldridgeBinaryPopulationSpectral2017} stellar population models to a composite rest-frame far-ultraviolet continuum spectrum of the VANDELS sources, following the method of \citet{cullenVANDELSSurveyStellar2019}.
Based on a stellar metallicity of $\simeq 0.07 \, \mathrm{Z}_{\odot}$, and assuming a 100 Myr constant star formation history, $\Rint$ for the individual galaxies varied between $0.17 - 0.20$ (depending on the redshift).

Here, we relax these assumptions and allow for greater variation of $\Rint$ by expanding the star formation timescales and stellar metallicities of the intrinsic spectrum.
We allow for stellar metallicities in the range $0.01 - 0.2 \, \mathrm{Z}_{\odot}$, constant star formation timescales $> 20$ Myr, and an upper-mass IMF cutoff between $100 \mathrm{M}_{\odot}$ and $300 \mathrm{M}_{\odot}$.
The resulting range of intrinsic ratios is $0.17 \leq \Rint \leq 0.38$.
We note that $\Rint$ can, in principle, be larger than $0.38$ at $< 20$ Myr, but we do not find strong evidence for such young ages in the individual VANDELS spectra.

\subsection{IGM and CGM Transmission}
\label{sub:trans}

At the redshift of our sample, it is the absorption of ionising photons by the intervening neutral hydrogen that has the largest influence on the derived value of the escape fraction, $\fesc$, and consequently its distribution. We see this in Equation \ref{eq:gen_model} as the complete degeneracy between the parameters $\fesc$ and $\trans$, which, mathematically, both denote a fraction that linearly reduces the value of the observed flux ratio, $\Robs$, compared to the intrinsic flux ratio $\Rint$.
Since we measure $\Robs$ with photometry, we are interested in the average fraction of produced ionising photons that are captured by our LyC filter, the U-band. We therefore let $\trans$ denote the average transmitted fraction of ionising photons, for a given sight line to a galaxy, integrated through the U-band filter (to remove the wavelength dependence).  

The optical depth, $\optdepth$, which determines the transmitted fraction of ionising photons within the wavelength range of the U-band, is dependent on the spatial distribution of neutral clouds, the HI column density of those clouds, and the redshift at which the photons are emitted. It will therefore vary significantly from one sight-line to another and between different galaxies. Since it is not possible to map the exact column densities and spatial distribution of neutral hydrogen clouds along the line of sight to each galaxy, we account for the transmitted fraction in a probabilistic manner. To account for both the IGM and CGM contribution to the transmission of LyC photons, we utilise transmission curves generated by \cite{begleyVANDELSSurveyMeasurement2022} using the parameterisation for the column density and redshift distribution of HI clouds given in \cite{steidelKeckLymanContinuum2018}. For six separate redshift bins ($z=3.4,3.5,3.6,3.7,3.8,3.9$), we generated 10,000 individual sight lines and integrated them through the U-band filter. For each redshift bin, we are thus provided with a distribution of the average transmission of LyC photons trough the LyC filter. We will utilise these distributions (see Figure 7 in \citealp{begleyVANDELSSurveyMeasurement2022}) as priors on the integrated transmitted fraction, $\trans$, in the Bayesian framework introduced in Section \ref{sec:bayes_inference}.

\subsection{UV Dust Attenuation}
\label{sub:auv}

After the effects of the IGM and CGM, the UV dust attenuation, $\Auv$, is the model parameter with the largest systematic influence on the escape fraction, $\fesc$. In this context, it is defined as the dust attenuation in the rest-frame effective wavelength of the V606 filter ($\lambda_\mathrm{eff,V606} = 5776.43\text{Å}$, corresponding to a typical rest-frame wavelength of $\simeq 1300\text{Å}$ for our final sample of galaxies \citep{begleyVANDELSSurveyMeasurement2022}.

The level of UV attenuation is determined individually for each galaxy, by comparing the observed UV spectral slope, $\beta_\mathrm{obs}$, with the spectral slope of the intrinsic SED model ($\beta_\mathrm{int}=-2.44\pm0.02$;  see Section \ref{sub:rint} and \citealp{begleyVANDELSSurveyMeasurement2022} for details of the fitting of an intrinsic model). The observed spectral slopes were determined by fitting a power law to each of the UV VANDELS spectra over the wavelength range $1300-1800\text{Å}$, within the continuum windows specified by \cite{calzettiDustExtinctionStellar1994}, obtaining an average of $\langle \beta_\mathrm{obs} \rangle = -1.26 \pm 0.04$ for the final galaxy sample. 
The attenuation at $\lambda=1600\text{Å}$ is determined from the spectral slopes, using the relation: $A_{1600} = 1.28 \cdot (\beta_\mathrm{obs}-\beta_\mathrm{int})$ \citep{begleyVANDELSSurveyMeasurement2022}. The conversion to the UV dust attenuation, $\Auv$, has a slight redshift dependence as the rest frame position of the effective wavelength of the V606 filter will vary depending on the redshift of the galaxy, but is roughly found by: $\Auv \simeq 1.2 \cdot A_{1600}$.

\section{Bayesian Inference}
\label{sec:bayes_inference}

To estimate the distribution of escape fractions across the population of galaxies in our sample, we adopt a two-level, hierarchical, Bayesian inference framework. With this hierarchical framework, we can combine information from individual galaxies to infer population-level parameters related to the distribution. 

In Section \ref{sub:single_layer}, we describe the mathematical framework of the individual-galaxy-level inference, while Section \ref{sub:hierarchical} details how the individual layers are combined hierarchically. Section \ref{sub:pop_dists} proposes four different, physically motivated parameterisations of the distribution of the escape fraction, and implements these into the general, hierarchical framework described in the previous section. Finally, Section \ref{sub:sample_configs} recounts the general configurations used for sampling the probability distribution with a Markov chain Monte Carlo (MCMC) algorithm.

Using Bayesian inference, we can incorporate prior knowledge in the framework, which becomes particularly useful when including the integrated transmission through the IGM and CGM. While we are unable to measure the true transmitted fraction of ionising photons for every line of sight, we do have the probability distributions of the integrated transmission in various redshift bins (see Section \ref{sub:trans}). The Bayesian framework allows us to draw upon this information in the form of a prior probability distribution.

\begin{figure}
    \centering


\definecolor{grey}{HTML}{DCDCDC}
\definecolor{faded}{HTML}{99A3A4}

\tikzset{
    > = latex, 
    every node/.append style = {
        draw = none,
        text = black
    },
    sampled_node/.style = {
        draw = black,
        shape = rectangle,
        rounded corners,
        fill = grey,
        inner sep = 13pt
    },
    diagnostic/.style = {
        draw = black,
        shape = rectangle,
        inner sep = 5pt
    },
    observed/.style = {
        draw = black,
        shape = circle,
        inner sep = 2pt
    },
    pdf/.style = {
        draw = black,
        dashed,
        shape = rectangle,
        inner sep = 5 pt
    }
}

\begin{tikzpicture}[scale=1] 


\node[sampled_node] (f) at (0,0) {};
\node[sampled_node] (Ri) at (1.9,0) {}; 
\node[sampled_node] (t) at (3,0) {};
\node[sampled_node] (A) at (4.1,0) {};
\node[sampled_node,scale=1] (phi) at (1,2) {};
\node[] (f_text) at (0,0) {$\fesci$};
\node[] (Ri_text) at (1.9,0) {$\Rinti$};
\node[] (t_text) at (3,0) {$\transi$};
\node[] (A_text) at (4.1,0) {$\Auvi$};
\node[scale=1.5] (phi_text) at (1,2) {$\pop$};

\node[pdf] (llh) at (1.7, -2.9) {$\mathcal{N}(\mathcal{R}_{\mathrm{model},i}, \sigma_{\mathcal{R},i}^2)$};

\node[] (sig_R) at (4, -2.9) {$\sigma_{\mathcal{R},i}$};

\node[diagnostic] (Rm) at (1.7, -1.8) {$\;\; \mathcal{R}_{\mathrm{model},i} \;\;$};

\node[observed] (Ro) at (1.7, -4.1) {$\Robsi$};

\path[->] (phi) edge (f);


\path[->] (0,-0.45) edge (Rm);
\path[->] (1.9,-0.3) edge (Rm);
\path[->] (3,-0.45) edge (Rm);
\path[->]  (4.1,-0.45) edge (Rm);

\path[->] (Rm) edge (llh);
\path[->] (sig_R) edge (llh);

\path[->] (llh) edge (Ro);


\path[-] (1.3,0.8) edge (5.3,0.8); 
\path[-] (1.3,-0.8) edge (5.3,-0.8); 
\path[-] (1.3,0.8) edge (1.3,-0.8); 
\path[-] (5.3,0.8) edge (5.3,-0.8); 
\node (Theta) at (5,0.5) {$\nuisi$}; 

\path[-,thick] (-1,1) edge (5.5,1); 
\path[-,thick] (-1,-4.7) edge (5.5,-4.7); 
\path[-,thick] (-1,-4.7) edge (-1,1); 
\path[-,thick] (5.5,-4.7) edge (5.5,1); 
\node (Local) at (4.4,-4.2) {Galaxy $i$}; 

\path[-,faded] (-0.9,1.1) edge (5.6,1.1); 
\path[-,faded] (5.5,-4.5) edge (5.6,-4.5); 
\path[-,faded] (5.6,-4.5) edge (5.6,1.1); 
\path[-,faded] (-0.9,1) edge (-0.9,1.1); 
\node (1st) at (0.9,0.9) {}; 
\path[->, faded] (phi) edge (1st);
\path[-,faded] (-0.8,1.2) edge (5.7,1.2); 
\path[-,faded] (5.6,-4.3) edge (5.7,-4.3); 
\path[-,faded] (5.7,-4.3) edge (5.7,1.2); 
\path[-,faded] (-0.8,1.1) edge (-0.8,1.2); 
\node (2nd) at (1.3,1) {}; 
\path[->, faded] (phi) edge (2nd);

\node (global) at (5,2) {Galaxy Population};

\end{tikzpicture}
    \caption{Graphical representation of the two levels in the hierarchical Bayesian inference framework. The method of inference for individual galaxy parameters (see Section \ref{sub:single_layer}) is contained within the main box (denoted Galaxy $i$), while the full figure includes the hierarchical extension to the Bayesian network incorporating the population parameters (see Section \ref{sub:hierarchical}). The grey outlines of multiple boxes in the background indicate the conditional independence between individual galaxy observations. Filled squares denote free, sampled parameters, while the rectangle with a solid border denotes a deterministic node (the value is unambiguously determined from the input parameters and Equation \ref{eq:gen_model}). The rectangle with the dashed border denotes a probability distribution with which the likelihood of the observed data, denoted with a circle, is evaluated. The observational error on $\Robs$, $\sigma_{\mathcal{R}}$ is a fixed input to the model.}
    \label{fig:diagram}
\end{figure}
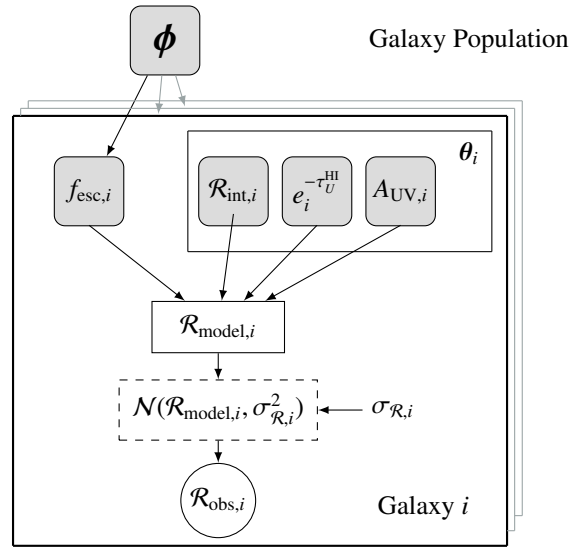

\subsection{Inference of Individual Galaxy Parameters}
\label{sub:single_layer}

At the individual galaxy level, we want to obtain the posterior distribution of the escape fraction, $\fesc$, given the observed data: the LyC to non-ionising UV flux ratio, $\Robs$. We can express the posterior probability of the escape fraction for one galaxy, $p(\fesc|\Robs)$, using Bayes' theorem:
\begin{equation}
    p(\fesc|\Robs) \propto \int p(\Robs|\fesc, \nuis) p(\fesc) p(\nuis) \; d\nuis,
    \label{eq:individual_posterior}
\end{equation}
where we have omitted the subscript $i$, which otherwise denotes each of the $N$ individual galaxies. Here, $\nuis = (\Rint, \trans, \Auv)$ is the set of additional model parameters needed in the generative model (see Equation \ref{eq:gen_model}). Since we focus on the escape fraction in this analysis, we refer to $\nuis$ as nuisance parameters and integrate over them in the expression above to obtain the marginal distribution. $p(\Robs|\fesc, \nuis)$ is the likelihood expressing how probable the observed data, $\Robs$, is, given the free parameters $\fesc$ and $\nuis$. $p(\fesc)$ is the prior on the escape fraction and $p(\nuis)=p(\Rint)p(\trans)p(\Auv)$ is the prior on the nuisance parameters.

We assume the likelihood function is normally distributed and  write the likelihood for a given galaxy as:
\begin{equation}
    p(\Robs|\fesc,\nuis) = \frac{1}{\sigma_R \sqrt{2\pi}} \exp \left( \frac{- (\mathcal{R}(\fesc,\nuis) - \Robs)^2}{2\sigma_R} \right),
\end{equation}
where $\sigma_R$ is the observation error on the flux ratio, $\Robs$, and $\mathcal{R}(\fesc,\nuis)$ denotes the modelled flux ratio as obtained with Equation \ref{eq:gen_model}.

When adopting a Bayesian approach, we furthermore need to specify prior distributions for the free parameters: the escape fraction, $\fesc$, and the 3 nuisance parameters contained in $\nuis$.
\begin{itemize}
    \item[$\bullet$] For the prior on the escape fraction, we define $p(\fesc)$ to be uniform between 0 and 1. 
    \item[$\bullet$] We assume a uniform prior on the intrinsic LyC to non-ionising UV flux ratio, $\Rint$ within a range that is redshift-dependent and determined from fitting stellar population models to the stacked VANDELS spectra (see Section \ref{sub:rint} for more details).
    \item[$\bullet$] For the integrated transmitted fraction, $\trans$, we use priors generated by \cite{begleyVANDELSSurveyMeasurement2022} from the distribution of sight lines described in Section \ref{sub:trans}. These are obtained by performing a bounded (to the domain $0\leq \trans \leq 1$) kernel density estimation (KDE)\footnote{The boundary correction of the KDE is made with the \texttt{linear\_combination} method (also known as generalised jackknifing) in \texttt{pyqt-fit} (\url{https://github.com/KimiKreil/pyqt-fit})} \citep{jonesSimpleBoundaryCorrection1993} of the distribution for each of the 6 redshift bins. For each galaxy, we use its spectroscopic redshift as a fixed input to determine which of the 6 numerical priors, $p(\trans)$, to use. The prior probability assigned to a sampled transmission is determined by interpolating the finite-resolution priors described here.
    \item[$\bullet$] The prior on the dust attenuation, $p(\Auv)$, is assumed to be a normal distribution with mean and standard deviation given by the fits to the UV continuum slope of each galaxy (See Section \ref{sub:auv}).
\end{itemize}
 
We provide a graphical representation of the inference setup in Figure \ref{fig:diagram}. The inference of individual galaxy parameters, described in this section, is displayed within the depicted main box (denoted "Galaxy $i$" in the illustration). Notice, however, that the prior on the escape fraction is predetermined for the inference of individual galaxy parameters which we are describing here.

\subsection{Hierarchical Inference of Population Parameters}
\label{sub:hierarchical}

This study aims to characterise the distribution of the escape fraction for our sample, $p(\fesc|\pop)$, where we have introduced a set of population parameters, $\pop$, to parameterise the distribution. At the population level, we want to obtain a posterior distribution of the set of parameters, $\pop$, given the observed data (i.e. the observed flux ratio measurements for the full sample of $N$ galaxies, $\{\Robs\}$). We keep the notation general in this section, reviewing the relevant equations with the use of a non-specified set of parameters, $\pop$, and defer the mathematical parameterisations of $\pop$ for four different model distributions to Section \ref{sub:pop_dists} along with their physical interpretation.

For one observation, i.e. one galaxy observation (corresponding to one box in Figure \ref{fig:diagram}), the posterior distribution of $\pop$ is described by:
\begin{equation}
    p(\pop|\Robsi) \propto p(\pop) \iint p(\Robsi|\fesci,\nuisi) p(\fesci|\pop)p(\nuisi) \; d\fesci d\nuisi, 
    \label{eq:pop_individual_post}
\end{equation}
where $i$ denotes the galaxy in question. $p(\pop)$ is the prior on the population parameters - which will be specified depending on the different parameterisations introduced in Section \ref{sub:pop_dists}. $p(\Robsi|\fesci, \nuisi)$ is the likelihood of the observed data given the free, local parameters ($\fesci$ and $\nuisi$) and our model (Equation \ref{eq:gen_model}). Finally, $p(\fesci|\pop)$ is the prior on the galaxy-individual escape fractions as determined by the population distribution described by the sampled $\pop$. Like the population prior, $p(\pop)$, the prior on the escape fraction is also dependent on the choice of model distribution, and we therefore refer to Section \ref{sub:pop_dists} for implementation details.

In Figure \ref{fig:diagram}, we visualise how the individual-galaxy-level and population-level parameters are connected in this hierarchical Bayesian inference framework. First, a set of population parameters, $\pop$, is sampled, which defines a prior distribution of escape fractions, from which we sample an escape fraction for each galaxy. Following the methodology for inferring individual galaxy parameters described in Section \ref{sub:single_layer}, we obtain samples of the posterior distribution of the population parameters as constrained by one galaxy: $p(\pop|\Robsi)$.

We can combine the inference from a set of conditionally independent observations (i.e. individual galaxies) by multiplying the individual posteriors (Equation \ref{eq:pop_individual_post}). The prior on the set of population parameters is, however, only applied once, since this term should only be invoked on the population level (i.e. outside the individual galaxy boxes in Figure \ref{fig:diagram}). The full expression for the posterior distribution of the population parameters, $\pop$, given the full set of $N$ observations, $\{\Robs\}$, thus becomes (after marginalising over nuisance parameters $\nuisi$):
\begin{equation}
    p(\pop|\{\Robs\}) \propto p(\pop) \prod_i^{N} \int p(\Robsi|\fesci) p(\fesci|\pop) \; d\fesci. 
    \label{eq:pop_posterior}
\end{equation}

We argue that we can replace the individual likelihoods, $p(\Robsi|\fesci)$, in Equation \ref{eq:pop_posterior} with the corresponding posterior distribution of the galaxy-specific escape fraction, $p(\fesci|\Robsi)$, from the individual galaxy inference, defined in Equation \ref{eq:individual_posterior} (see Section \ref{sub:single_layer}). The individual likelihoods we are interested in for Equation \ref{eq:pop_posterior} are defined as (marginalised over the nuisance parameters $\nuisi$):
\begin{equation}
    p(\Robsi|\fesci) = \frac{p(\fesci|\Robsi)p(\Robsi)}{p(\fesci)}.
\end{equation}
For the inference of individual galaxy parameters, the prior on $\fesci$ is uniform and therefore corresponds to scaling $p(\fesci|\Robsi)$ by a constant $p(\fesci)$. The evidence, $p(\Robsi)$, is constant for the choice of model (Equation \ref{eq:gen_model}), and will therefore, similarly, only result in a scaling of $p(\fesci|\Robsi)$. This implies that $p(\Robsi|\fesci) \propto p(\fesci|\Robsi)$, and the proposed replacement is therefore valid due to the normalisation-insensitive property of an MCMC algorithm. Note that we only obtain samples of the posterior distribution with an MCMC algorithm, and we therefore have to perform a KDE on the set of posterior samples for each $p(\fesci|\Robsi)$ to efficiently use the individual posteriors as likelihoods in the hierarchical framework.

\subsection{Population Distributions}
\label{sub:pop_dists}

\begin{figure*}
    \centering
    \includegraphics[width=\textwidth]{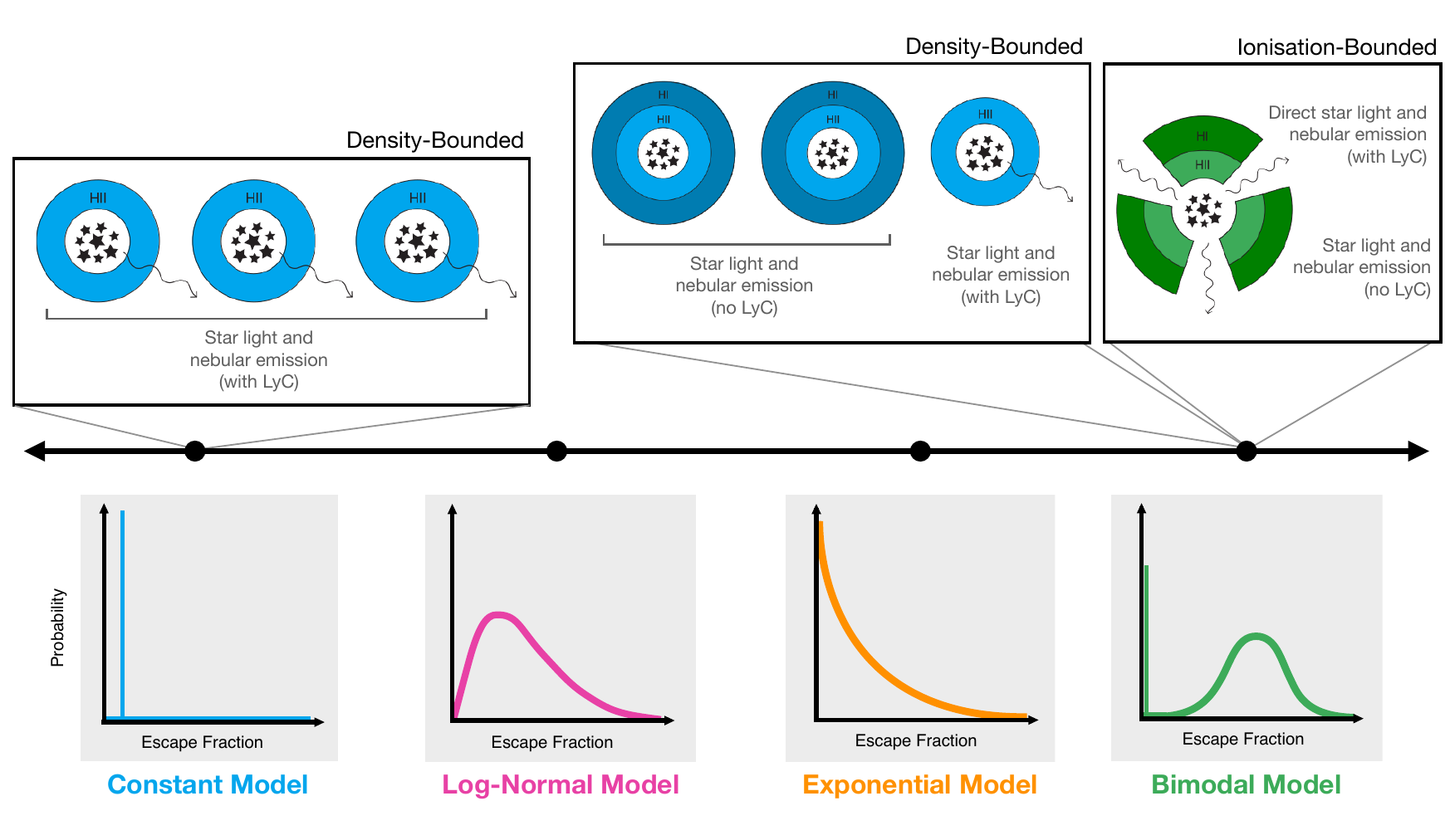}
    \caption{The four physically motivated distributions of the escape fraction placed on a scale to indicate their connection to the two simplified physical pictures of how ionising photons may escape the ISM of their host galaxy. The "blue picture" represents the density-bounded leakage mechanisms, while the "green scenario" is where ionising photons escape through ionised, low-density channels. In both scenarios, arrows indicate the escape of LyC photons. The constant distribution represents a scenario where all galaxies have the same physical conditions and LyC photons escape via purely density-bounded leakage from Strömgren spheres with the same column density, while the bimodal model represents either a purely ionisation-bounded scenario where leakage is strongly sightline dependent and/or strongly time variable. The log-normal and exponential distribution represent complex mixtures of the two main leakage mechanisms.}
    \label{fig:physical_picture}
\end{figure*}

In this section, we introduce four different parameterisations for the population distribution of escape fractions, namely, (i) a constant escape fraction across the population of galaxies (following \citealp{begleyVANDELSSurveyMeasurement2022}), (ii) a lognormal distribution (iii) an exponentially decaying distribution, and (iv) a bimodal distribution consisting of a constant and a normal component. 

Here, we specify the set of population parameters, $\pop$, the prior on the population parameters, $p(\pop)$, and provide the functional form of the population distribution (which assumes the purpose of a prior), $p(\fesci|\pop)$, for the four different model distributions. The distributions are all physically motivated and in Figure \ref{fig:physical_picture} we demonstrate each is connected to the two main mechanisms with which ionising photons escape their host galaxy: a density-bounded and/or an ionisation-bounded scenario.

Common for the four parameterisations, is that the escape fraction, $\fesc$, is defined to be between 0 and 1. The distribution of escape fractions, $p(\fesci|\pop)$, must therefore be normalised in this interval. For simplicity, we omit this normalisation of the truncated distribution in the notation for the four different distributions described in the subsequent sections.

\subsubsection{Constant Distribution}

This model assumes that the escape fraction, $\fesc$, is the same for the entire galaxy population. The distribution of escape fractions, $p(\fesc|\pop)$, will thus be described by a Dirac-delta function placed at $\fesc=\mu$. The set of population parameters for this model is thus given by a single parameter, $\pop_\mathrm{const}=\mu$, and the distribution takes the functional form:
\begin{equation}
    p(\fesc|\pop_\mathrm{const}) = \delta(\fesc-\mu),
    \label{eq:const}
\end{equation}
where $\delta$ denotes the Dirac-delta function. We assume a uniform prior between 0 and 1 on $\mu$.

The constant distribution represents the case where the fraction of LyC photons escaping is the same for different galaxies such that a single value of the escape fraction is descriptive of all observations in our sample. In this scenario the leakage of ionising photons is exclusively driven by a density-bounded scenario, assuming all galaxies have Strömgren spheres of the same density. In the density-bounded scenario, surrounding hydrogen has been ionised by LyC flux from a powerful star-formation period in the centre of the galaxy. In this picture, a fraction of the subsequently produced ionising photons is used to balance the rate of recombination, while the remaining fraction of LyC photons is allowed to escape the galaxy's ISM.

\subsubsection{Log-Normal Distribution}

Here, the population distribution of escape fractions, $p(\fesc|\pop)$, is given by a log-normal distribution, parameterised by:
\begin{equation}
    p(\fesc|\pop_\mathrm{lognorm}) \propto \frac{1}{\fesc \sigma \sqrt{2\pi}} \cdot \exp \left(-\frac{(\ln(\fesc)-\mu)^2}{2\sigma^2} \right),
\end{equation}
where the population parameters are $\pop_\mathrm{lognorm}=(\mu, \sigma)$, which describe, respectively, the mean and standard deviation of the natural logarithm of the variable $\fesc$. 

To retain the characteristic feature of a log-normal distribution within the domain of the escape fraction, we require that the mode, $M=\exp(\mu-\sigma^2)$, is always between 0 and 1. This implies that the prior on the standard deviation should restrict $\sigma$ to the range as determined by the sampled $\mu$ and inserting $M=0$ and $M=1$ in the expression: $\sigma=\sqrt{\mu-\log(M)}$. By definition, we need $\sigma>0$ and we therefore require $\mu-\log(M)>0$ and $M>0$. Furthermore, to avoid computational overflow we limit $\sigma>0.01$ and for simplicity assume $0<\mu<5$. We employ uniform priors for both parameters in the specified ranges. In practice, we sample $\log_{10}(\sigma)$ to more effectively explore the parameter space.

The log-normal distribution represents a complex mixture of leakage mechanisms. In particular, the distribution of escape fractions in this physical picture could arise from an underlying log-normal distribution of neutral hydrogen column densities in the galaxy's ISM. In this distribution, the escape fraction cannot be zero, essentially ruling out a purely ionisation-bounded scenario or strongly time-variable leakage.

\subsubsection{Exponential Distribution}

We assume the population distribution of escape fractions, $p(\fesc|\pop)$, is given by an exponentially decaying function, parameterised as:
\begin{equation}
    p(\fesc|\pop_\mathrm{exp}) \propto \frac{1}{\mu} e^{- \fesc \; / \; \mu},
\end{equation}
where $\mu>0$ is the scale parameter, which also describes the mean of the distribution. We require the mean escape fraction to be between $0$ and $1$ and therefore assume a uniform prior on this interval. In practice, we sample the parameter in logarithmic space and thus enforce a uniform prior between -3 and 0 on the transformed parameter $\log_{10}(\mu)$. 

Similar to the log-normal distribution, the exponential distribution represents a complex mixture of leakage mechanisms and a potentially exponential distribution of HI column densities. $\fesc =0$ is possible in this model, allowing for strongly time- and/or sightline-variable LyC leakage.

\subsubsection{Bimodal Distribution}

In the bimodal model distribution, we assume the population distribution of escape fractions, $p(\fesc|\pop)$, is made up of two components: (i) a fraction, $w$, of the galaxies in the population is assumed to exhibit a zero escape fraction, i.e. no LyC photons escape from these galaxies. This component is described by a Dirac-delta function placed at $\fesc=0$, and weighted by $w$. (ii) the escape fraction of the remaining galaxies in the population is described by a normal distribution, centred at $\mu$ with standard deviation $\sigma$. The set of population parameters for this parameterisation is thus given by $\pop_\mathrm{bimodal}=(w, \mu, \sigma)$, and the population distribution is described by:
\begin{equation}
    p(\fesc|\pop_\mathrm{bimodal}) \propto w \cdot \delta(\fesc) + (1-w) \cdot \mathcal{N}(\fesc|\mu, \sigma^2),
\end{equation}
where $\delta$ denotes the Dirac-delta function and $\mathcal{N}$ denotes the normal distribution.

The prior on the population parameters is given by: $p(\pop)=p(w)p(\mu)p(\sigma)$. We assume the prior on the weight parameter, $w$, is uniform between 0 and 1 such that the full distribution of escape fractions normalises to unity (after truncating the normal distribution component). Furthermore, we assume the mean, $\mu$, of the normal distribution must be part of the domain of the escape fraction, and we thus assume the prior to be uniform between 0 and 1 as well. Lastly, the standard deviation of the normal distribution, $\sigma$, should be positive. For computational efficiency, we restrict this parameter more and sample it in logarithmic space. We use a uniform prior on the interval $-3 \leq \log_{10}(\sigma) \leq 0$.

The two components of the bimodal distribution represent, respectively, a fraction of galaxies with zero escape fraction corresponding to galaxies where no LyC photons escape and a fraction of galaxies with higher escape fractions corresponding to observations where the line of sight coincides with a low-density channel in the surrounding hydrogen. As such, the bimodal distribution represents either an ionisation-bounded scenario with a small number of highly ionised, low-density channels, and/or strongly time variable star formation and LyC escape. In the latter case, galaxies are optically thin to LyC radiation for only a short timespan, for example, due to hard radiation fields and/or stellar or supernovae feedback, and optically thick the rest of the time.

\subsection{Sampling Configurations}
\label{sub:sample_configs}

To sample the various posterior distributions of interest, we employ the affine-invariant, ensemble, Markov chain Monte Carlo (MCMC) sampler \texttt{emcee} \citep{foreman-mackeyEmceeMCMCHammer2013} with the default \texttt{StretchMove}, a \texttt{Python} implementation of the sampling algorithm originally proposed by \citep{goodmanEnsembleSamplersAffine2010}. 

The walkers are initialised with positions in the parameter space drawn from the respective parameter priors, except for the integrated transmission values, which are initialised uniformly between 0 and 0.5 (instead of the full range spanning up to 1). The transmission priors were generated with a KDE of 10,000 sight line simulations, but for transmitted fractions closer to 1 we only have a few simulated sightlines and we enter a regime of low statistics (especially for higher redshifts). As a result, some of these priors exhibit local extrema that are unphysical artefacts resulting from low sampling. To avoid walkers getting stuck in a local, unphysical maximum we initialise the walkers closer to the true maxima found between 0 and 0.5.

We use 50 walkers, each sampling at least 10,000 positions in the parameter space. The number of steps is chosen so the chains are at least 100 times the longest integrated auto-correlation length, thus ensuring independent samples of all the free parameters. The number of samples discarded as the burn-in phase is computed as $2\cdot \max(\tau)$, where $\tau$ is a list of auto-correlation lengths for each of the sampled parameters, and we use every $0.5\cdot \min(\tau)$ step of the sampled chains.

Unless otherwise stated, we report inferred model parameters as the mode of the corresponding posterior samples along with the 68th per cent highest posterior density (HDP) interval. 
\section{Results}
\label{sec:results}

In this section, we present our constraints on the distribution of escape fractions estimated from the observed flux ratios, $\{\Robs\}$, from a sample of 148 star-forming, EoR-analogues ($z \simeq 3.5$) galaxies in the VANDELS survey \citep{mclureVANDELSESOPublic2018,pentericciVANDELSESOPublic2018,garilliVANDELSESOPublic2021}.

Section \ref{sub:results_single} presents the constraints on the escape fraction, $\fesc$, and the nuisance parameters, $\nuis$, for individual galaxies, while Section \ref{sub:results_pop} describes the inferred constraints on the population distribution of escape fractions, $p(\fesc|\pop)$, based on observations from all galaxies in the sample. Here, we provide constraints for the four different parameterisations of the distribution: constant, log-normal, exponential and bimodal (all described in Section \ref{sub:pop_dists}). Finally in Section \ref{sub:results_comparison}, we compare the four model distributions and investigate their respective ability to reproduce the observed distribution of $\Robs$ in the galaxy sample.

\subsection{Constraints on Galaxy Specific Escape Fractions}
\label{sub:results_single}

\begin{figure}
    \centering
    \includegraphics[width=\textwidth/2]{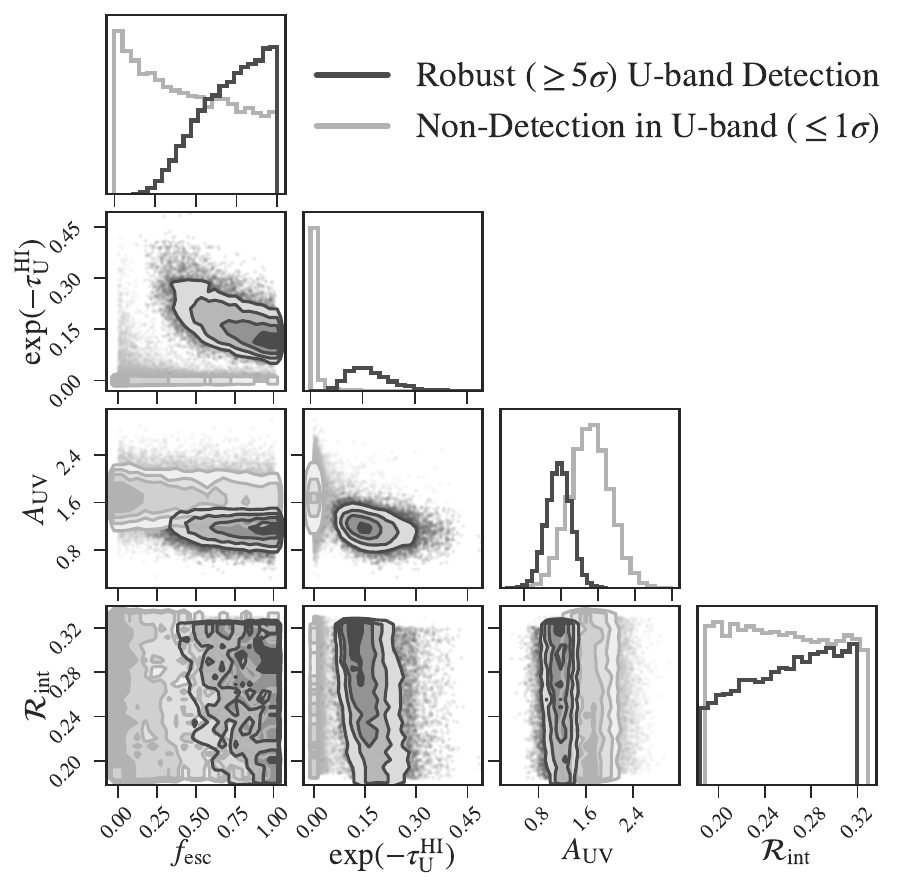}
    \caption{Corner plot showing the 1D and 2D marginalised posteriors for the escape fraction, $\fesc$, the integrated transmission through the U-band filter, $\trans$, the dust attenuation at the effective wavelength of the V606 filter, $\Auv$, and the intrinsic LyC to non-ionising UV flux ratio, $\Rint$, for two example galaxies in our sample. In light grey, we show the posterior distributions for a typical non-detection in the U-band filter ($\leq 1\sigma$). In dark grey, we show the posterior distributions for a robust $(\geq 5\sigma)$ $\Robs$ detection in the U-band. The posteriors are obtained using an inference scheme incorporating only individual galaxy parameters, as described in Section \ref{sub:single_layer}.}
    \label{fig:single_corner}
\end{figure}

We use the framework for inferring individual galaxy parameters (see Section \ref{sub:single_layer}) to constrain the escape fraction, $\fesc$, for each of the 148 galaxies in our sample. In Figure \ref{fig:single_corner}, we show examples of the resulting posterior distributions (not explicitly marginalised over the nuisance parameters) for a robust detection ($\geq 5 \sigma$) and a typical, non-detected galaxy in the U-band ($\leq 1 \sigma$).

Inspecting the marginalised posterior of the escape fraction (see the top panel of Figure \ref{fig:single_corner}), we see that even for the galaxy with a robust $\Robs$ detection (dark grey), we are not able to tightly constrain the escape fraction. The posterior provides a lower limit on the escape fraction ($\fesc>0.38$ from the 68 per cent HDP interval), while the case is even less informative for a typical non-detection (light grey), where the posterior samples the full domain of the escape fraction. These results indicate that observations from single galaxies do not individually exhibit much constraining power on the galaxy-individual escape fractions. In the next section, however, we will demonstrate how the constraining power on population-level parameters profits from combining observations of multiple galaxies.

\subsection{Constraints on the Population Distribution of Escape Fractions}
\label{sub:results_pop}

\begin{figure*}[h!]
  \centering
  \subfloat[][Constant Model]{\includegraphics[width=.45\textwidth, height=.45\textwidth]{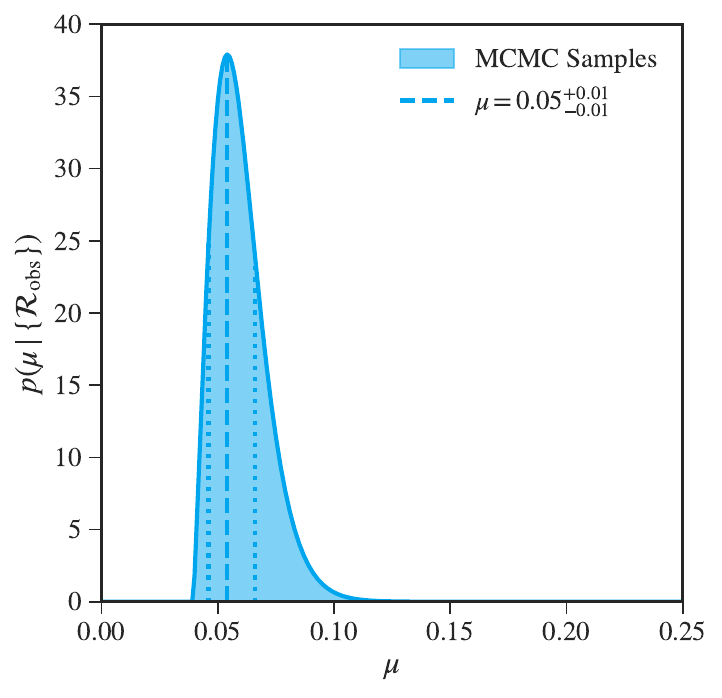}}\quad
  \subfloat[][Log-Normal Model]{\includegraphics[width=.45\textwidth, height=.45\textwidth]{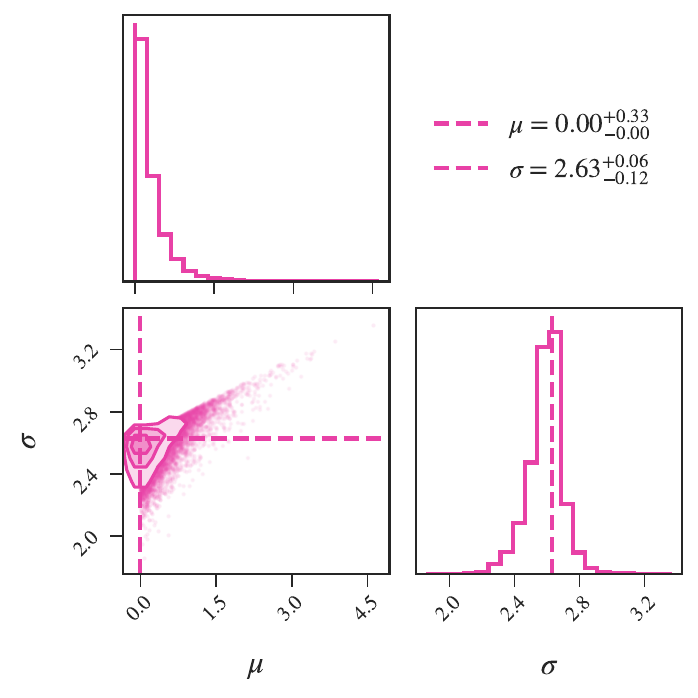}}\\
  \subfloat[][Exponential Model]{\includegraphics[width=.445\textwidth, height=.445\textwidth]{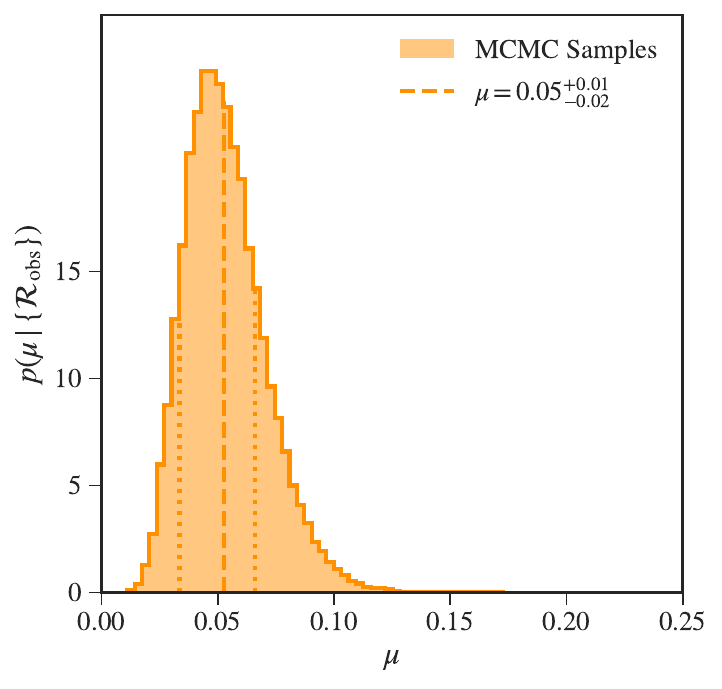}}\quad
  \subfloat[][Bimodal Model]{\includegraphics[width=.442\textwidth, height=.442\textwidth]{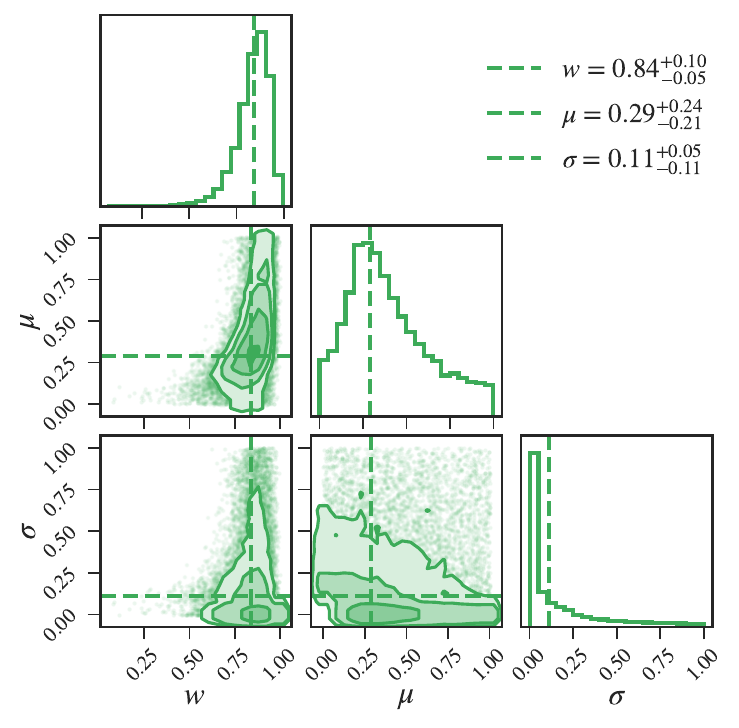}}
  \caption{1D and 2D marginal posterior distributions of the set of population parameters, $\pop$, for each of the four models described in Section \ref{sub:pop_dists}. Dashed lines indicate the mode of the posterior samples, and values are reported as the mode with the 68 per cent HDP interval. \textbf{Panel (a)} displays the 1D marginal posterior of the population parameter $\pop_\mathrm{const}=\mu$ (blue), describing the constant escape fraction value most likely to describe all galaxies in the sample. \textbf{Panel (b)} provides a corner plot showing the 1D and 2D marginal posterior distributions of the set of population parameters $\pop_\mathrm{lognorm}$ (pink), describing a log-normal distribution of escape fractions. Here $\mu$ and $\sigma$ represent, respectively, the mean and standard deviation of the natural logarithm of the escape fraction. \textbf{Panel (c)} shows the 1D marginal posterior of the population parameter $\pop_\mathrm{exp}=\mu$ (orange), which represents the scale parameter in an exponential distribution of escape fractions. Finally, \textbf{panel (d)} shows the corner plot containing the 1D and 2D marginal posteriors for the set of population parameters $\pop_\mathrm{bimodal}$ (green), which describes a bimodal distribution of escape fractions: $w$ denotes the fraction of galaxies with a constant escape fraction equal to zero, while $\mu$ and $\sigma$ describe the escape fraction of the remaining galaxies as being normally distributed.}
  \label{fig:model_posteriors}
\end{figure*}

We use the general, hierarchical inference framework, described in Section \ref{sub:hierarchical}, to constrain the set of population parameters, $\pop$, for each of the four parameterisations of the escape fraction distribution: (a) constant, (b) log-normal, (c) exponential and (d) bimodal (all described in Section \ref{sub:pop_dists}). In Figure \ref{fig:model_posteriors}, we display the resulting posterior distributions for the set of population parameters, $\pop$, for each of the four models.

Panel (a) (of Figure \ref{fig:model_posteriors}) shows the marginalised posterior distribution of the constant distribution's population parameter, for which we report an inferred value of $\ConstAll$. This distribution assumes that all galaxies share the same value of the escape fraction and $\mu$ should therefore be interpreted as the most likely value of $\fesc$ to describe all galaxies, rather than the average of 148 individual values of the escape fraction. Our inferred value is fully consistent with the corresponding $\langle \fesc \rangle = \BegleyBayesfesc$ that was reported by \cite{begleyVANDELSSurveyMeasurement2022}, using a similar approach.

We note here there are two key differences between our implementation and that of \citet{begleyVANDELSSurveyMeasurement2022}, but both effects effectively cancel out to provide the same $\langle \fesc \rangle$ result. 
Firstly, by including a variable intrinsic flux ratio, $\Rint$ in our inference, we are allowing higher values of $\Rint$ compared to \citet{begleyVANDELSSurveyMeasurement2022} which translates to slightly lower values of the escape fraction, $\fesc$.
Secondly, we update the KDE method used to obtain the individual $\fesc$ posteriors so they can be multiplied, obtaining the population posterior displayed in Figure \ref{fig:model_posteriors}. 
We use a KDE bounded at $0 \leq \fesc \leq 1$, which is more robust for high $\fesc$ and cause a systematic shift to slightly higher $\fesc$, contrary to \cite{begleyVANDELSSurveyMeasurement2022} who used only a lower bound on the KDE. Replicating the method in \cite{begleyVANDELSSurveyMeasurement2022} where $\Rint$ is a fixed parameter, but with our updated KDE implementation, we find a slightly higher $\langle \fesc \rangle = 0.07^{+0.02}_{-0.01}$ compared to than presented by \citet{begleyVANDELSSurveyMeasurement2022}.

Panel (b) (of Figure \ref{fig:model_posteriors}) shows the 1D and 2D marginalised posterior distributions for the population parameters constituting the log-normal distribution. The set of population parameters is found to be $\LognormAll$, describing, respectively, the mean and the standard deviation of the natural logarithm of the escape fraction. The expected value is $E[\fesc]=0.29$, i.e. higher than for the constant distribution, because the shape of the log-normal distribution does not allow for $\fesc=0$, and thus produces nonphysical results as we will discuss in Section \ref{sub:results_comparison}.

Panel (c) (of Figure \ref{fig:model_posteriors}) shows the marginal posterior distribution of the scale parameter of the exponential model distribution. We find that $\ExpAll$. The scale parameter, $\mu$, also describes the mean escape fraction of the population distribution, $p(\fesc|\pop_\mathrm{exp})$. The expected value is $E[\fesc]=0.05$ and is thus consistent with the constant model. 

Panel (d) (of Figure \ref{fig:model_posteriors}) shows the 1D and 2D marginalised posterior distributions for the population parameters in the bimodal parameterisation. The set of population parameters is found to be $\BimodalAll$, denoting, respectively the fraction of galaxies with a constant zero escape fraction, and the mean and standard deviation of the normal component. The expectation value is $E[\fesc]=0.05$, fully consistent with the constant and exponential model parameterisations.

While the expectation values of the four inferred escape fraction distributions mostly align well, this does not provide insight into the actual shapes of the distributions and how they may differ. 

\begin{figure}
    \centering
    \includegraphics[width=\textwidth/2]{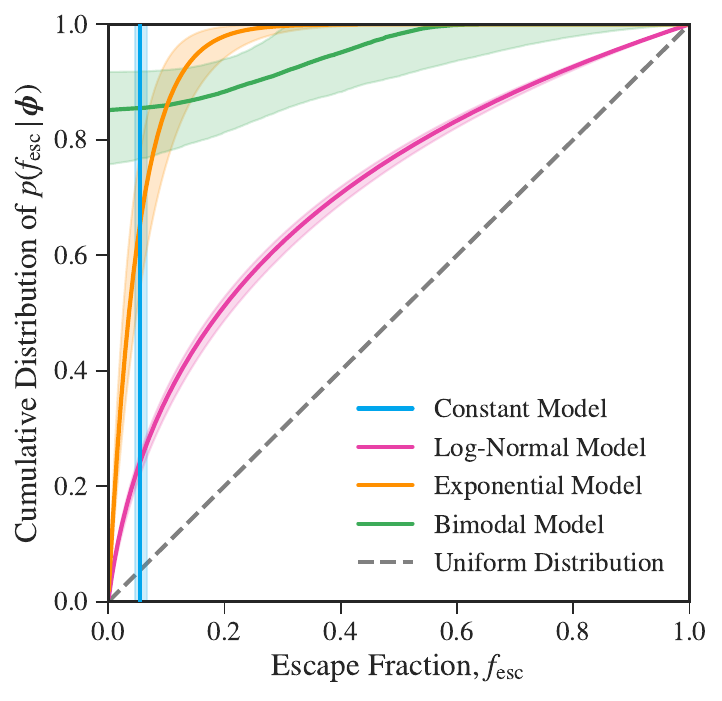}
    \caption{Median cumulative distribution of the population distribution of escape fractions for each of the four models. To compute this, for each model we: (i) select \Nexp parameter-sets from the chains of posterior samples displayed in Figure \ref{fig:model_posteriors}, (ii) evaluate the CDF of the distribution given by the selected population parameters on a grid of escape fractions, (iii) for each index of the escape fraction grid, compute the median of the 10,000 corresponding CDF values. The median is shown in a solid line, while the shaded region marks the 16th and 84th percentile intervals. For reference, we plot the CDF of a uniform distribution in the dashed, grey line.}
    \label{fig:CDF_comparison}
\end{figure}

To visually compare the different models, in Figure \ref{fig:CDF_comparison}, we plot the median cumulative distribution function (CDF) of the escape fractions for each of the four models. The median CDF, for each model, is computed from the resulting distributions of \Nexp parameter-sets taken from the posterior samples\footnote{Except for the constant model which by definition will have a strictly vertical CDF and is therefore plotted using the corresponding quantiles of the sampled parameter $\phi_\mathrm{const}=\mu$} displayed in Figure \ref{fig:model_posteriors}. The median is shown in a solid line, while the shaded region marks the 16th and 84th percentile intervals. 

The constant, exponential and bimodal models all indicate that the majority (or all) of the galaxies in our sample exhibit low escape fractions ($\fesc<0.1$), which agrees well with the large number of non-detections in our sample (144 galaxies with $\text{S}/\text{N}<3$). The log-normal distribution similarly predicts a majority of galaxies with lower escape fractions (50\% of galaxies with $\fesc<0.2$), but predicts more galaxies with larger escape fractions compared to the other distributions.  

\subsection{Model Comparison}
\label{sub:results_comparison}

We now aim to constrain which of the four distributions best describes our observations.
Adopting an MCMC approach to infer the distribution of escape fractions, results presented in the previous section are inherently dependent on the chosen parameterisation of the distribution.
To properly quantify whether an inferred distribution is capable of describing the observed data (the LyC to non-ionising UV flux ratios of the galaxies, $\{\Robs\}$), we perform a posterior predictive check \citep[e.g.,][]{Gelman1996}. We produce a mock $\Robs$ distribution based on the inferred escape fraction distribution and compare it to the observed data using a two-sample Kolmogorov-Smirnov (KS) test. 

The KS test statistic is a measure of how much the two distributions differ (it reports the maximum difference between the two cumulative distributions) and a lower KS score will therefore indicate a better model. Together, the KS score and the sample size translates into a probability (p-value) of whether the two $\Robs$ distributions are samples from an identical population distribution. The significance level, $\alpha$, corresponds to the desired, minimum probability, i.e. when we set $\alpha=0.10$, we reject a model if a resulting mock $\Robs$ distribution exhibits less than a 10\% probability of matching the observed distribution. Since the KS test is a non-parametric test, the p-value tests for any violation of the null-hypothesis, for example if the $\Robs$ distributions have different medians, different variances or different population distributions. It thus provides a flexible, yet powerful test that is furthermore independent of any binning of the data.

To produce the mock $\Robs$ distribution for each of our four models, we first sample a set of population parameters from the posterior chains (displayed in Figure \ref{fig:model_posteriors}) each defining a certain distribution of escape fractions. From this distribution, we sample 148 escape fractions to produce a sample of mock galaxies with the same size as our true sample. To best emulate the real sample, we use the (for each galaxy) measured spectroscopic redshift, UV dust attenuation, $\Auv$, along with the observed LyC to non-ionising UV flux ratio, $\Robs$, (and error $\sigma_\mathcal{R}$) as parameters for creating a mock sample. The fitted intrinsic SED (described in Section \ref{sub:rint}) is then modified to produce an observed spectrum for each galaxy and $\Robs$ is extracted (see Appendix A for the full procedure of producing mock galaxy spectra and measuring $\Robs$). This process is repeated \Nexp times, essentially producing \Nexp mock $\Robs$ distributions, for each of the four models presented in Section \ref{sub:pop_dists}. These many distribution samples are then collapsed into one distribution with $\Nexp \times 148$ samples, so we can perform a single KS test for each model and thus benefit from the self-normalising property of the KS test.

With such a large sample size, the sampling error is negligible and we choose to employ a somewhat large  $\alpha=10\%$ significance level (compared to the conventional 5\%) for rejecting the hypotheses of the distribution matching the observed data.

\begin{figure}
    \centering
    \includegraphics[width=\textwidth/2]{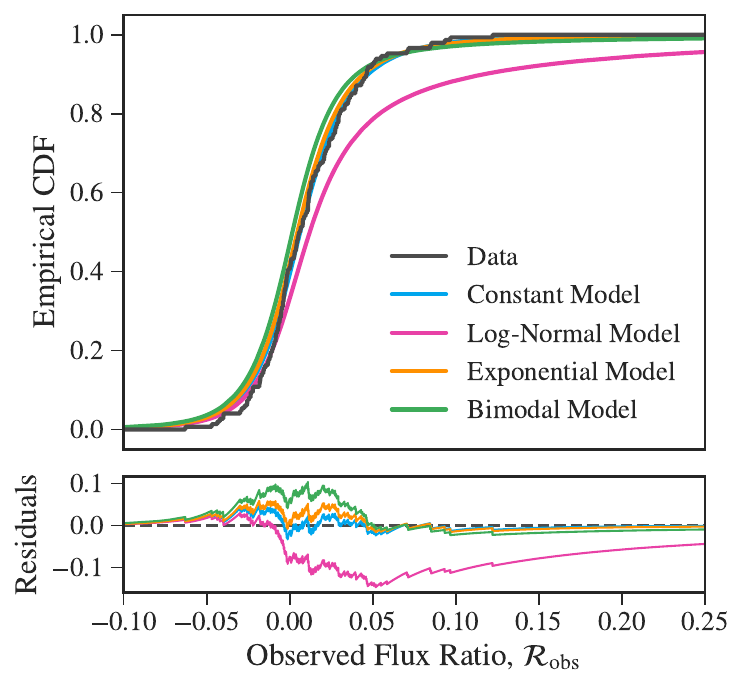}
    \caption{The top panel provides a visual illustration of the KS score, which measures the maximum difference between the eCDF of the observed flux ratios, $\Robs$ (plotted in black, top panel) and the eCDF of a simulated $\Robs$ distribution dependent on respectively the constant, log-normal, exponential and bimodal model distribution of the escape fraction as inferred from the Bayesian framework. The bottom panel shows the residual (the model eCDF minus the data eCDF) to highlight the difference between the two, i.e. what the KS test measures.}
    \label{fig:KS}
\end{figure}

\begin{table}[]
\centering
\begin{tabular}{lll}
\hline
Model Distribution & KS Test Statistic & p-Value \\ \hline
Constant           & 0.048             & 0.866   \\
Log-Normal          & 0.147             & 0.003   \\
Exponential        & 0.059             & 0.658   \\
Bimodal            & 0.104             & 0.076   \\ \hline
\end{tabular}
\caption{The test statistic and the associated p-value calculated with the Kolmogorov Smirnov test, comparing the distribution of the observed flux ratio, ${\Robs}$, to mock $\Robs$ distributions produced with samples of the escape fraction drawn from respectively the constant, log-normal, exponential and bimodal distribution of the escape fraction inferred in this work.}
\label{tab:ks}
\end{table}

In Figure \ref{fig:KS}, we display the empirical CDF (eCDF) of the real, observed sample of flux ratios, ${\Robs}$, together with the corresponding eCDF for each of the four model distributions. Here, the KS score for each model represents the largest vertical distance between the two. From visual inspection it is evident that the log-normal model performs worst. It significantly overestimates the number of galaxies with observed flux ratios above $\Robs > 0$, which is related to this model predicting more galaxies with larger escape fractions compared to the other models, as we saw in Figure \ref{fig:CDF_comparison}. Furthermore, we see that the bimodal model slightly overestimates the number of non-detections with $\Robs$ around 0.

We provide the KS test score and the associated p-value for each of the four model distributions in Table \ref{tab:ks}. At a 10\% significance level, we reject the log-normal and bimodal $\fesc$ distributions as a good description of our observed flux ratios, ${\Robs}$, while both the constant and exponential models exhibit plausible p-values. However, as we argue in Section \ref{sec:discussion}, the exponential model is likely the best representative of the escape fractions of our galaxy sample.

\subsection{Correlation with the UV Continuum Slope}
\label{sub:beta-split}

We investigate how the escape fraction correlates with the UV continuum slope, $\beta$ (a proxy for dust and HI column density of the ISM in the galaxy), for our best-fitting model: the exponential distribution. 

We split the full galaxy sample in half at the median value of the UV slope, $\beta=-1.30$, creating two subsamples (red and blue galaxies) each consisting of 74 galaxies. For each of these samples, we use the hierarchical Bayesian inference framework to determine the population distribution of escape fractions as parameterised by the exponential distribution.

\begin{figure}
    \centering
    \includegraphics[width=\textwidth/2]{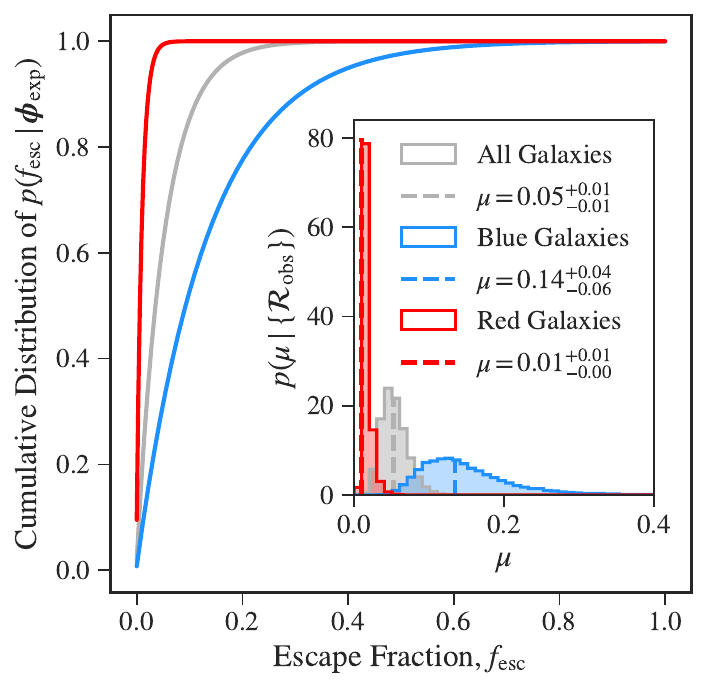}
    \caption{Cumulative distribution of the escape fraction as parameterised by the exponential scale parameter, $\pop_\mathrm{exp}=\mu$, inferred for, respectively, the full sample (grey), the galaxy subsample with lowest UV slopes (blue) and the subsample with highest UV slopes (red). The inserted panel shows the corresponding inferred posterior distribution, where the dashed line marks the mode of the distribution.}
    \label{fig:beta-split}
\end{figure}

Figure \ref{fig:beta-split} shows the resulting CDF of the escape fraction distribution for, respectively, the full sample, the red subsample, and the blue subsample. We find that there is a correlation between higher escape fractions and lower UV slope (bluer galaxies), indicating that galaxies with lower levels of UV dust attenuation display a larger escape fraction.

\section{Discussion}
\label{sec:discussion}

In Section \ref{sub:best_model}, we discuss which escape fraction distribution is best representative of our sample and what implications it has for our understanding of how ionising photons escape their host galaxy. We compare our best model distribution with predictions from simulations in Section \ref{sub:sim_comparison}. For a discussion on systematic uncertainties associated with modelling the escape fraction with Equation \ref{eq:gen_model}, we refer to Section 5 in \cite{begleyVANDELSSurveyMeasurement2022}.

\subsection{What is the Best Representation of the Population Distribution of the Escape Fraction?}
\label{sub:best_model}

This work presents the first constraints on the shape of the escape fraction distribution in the galaxy population at $z\sim3-4$. Previous work by \citet{vanzellaGREATOBSERVATORIESORIGINS2010} had measured median values given different escape fraction distributions but did not distinguish between the models. Here we have taken the important next step in performing a model comparison to assess which model best fits the observations.

The constant and exponential model of the escape fraction distribution both exhibit acceptable p-values when employing a KS-test and we therefore cannot reject either of their ability to match the observed data. Here, we argue that, physically, the ionising photon escape fraction should not be a constant across the galaxy population and thus the exponential model should be the best description of the data.

Our sample of galaxies includes 2 objects with robust LyC detections (individual U-band flux detections with $\geq 5 \sigma$ significance), for one of which we presented the constraints on the galaxy-individual escape fraction in Figure \ref{fig:single_corner}. The 95 per cent highest posterior density interval places a lower limit on the escape fraction for this galaxy at $\fesc>0.38$, clearly incompatible with the physical picture proposed by the constant model, where we inferred a constant escape fraction of $\mu=\ConstMuVal$ (See panel (a), Figure \ref{fig:model_posteriors}). Furthermore, these 2 objects have previously been reported in the literature \citep[e.g.,][]{vanzellaGREATOBSERVATORIESORIGINS2010, jiHSTImagingIonizing2020, saxenaNoStrongDependence2022} which finds similar lower limits on the galactic scale escape fractions that conflict with the scenario described by the constant model. Therefore, we argue, that the constant distribution cannot be the most representative model. 

Our inferred, exponential distribution of the escape fraction points towards complex LyC photon leakage scenarios, where most galaxies exhibit similar, non-zero, but low escape fractions ($\fesc<0.1$) while a few galaxies exhibit high leakage of LyC photons. This may reflect a broad distribution of ISM HI column densities and strongly time and/or sightline variable LyC leakage where for a high fraction of time and/or sightlines galaxies are optically thick to LyC photons.

Our sample only presents a few actively LyC leaking galaxies, explaining why a constant model can statistically reproduce the data and why the parameters describing the normal component of the bimodal model, $\pop_\mathrm{bimodal}=(\mu,\sigma)$, are difficult to constrain (See Figure \ref{sub:results_pop}, panel (d)). We find support for the exponential model being the best representation of the population distribution of the escape fraction in our sample, but the analysis would benefit from a larger sample to better constrain the respective model parameters - in particular, to better capture the tail of the distribution representing LyC leakers.

\subsection{Comparison with Simulations}
\label{sub:sim_comparison}

To get physical insight into the inferred escape fraction distribution we can compare to predictions from simulations. There exist multiple frameworks for simulating reionisation, all reporting some distribution of the escape fraction \citep[e.g.,][]{kimmESCAPEFRACTIONIONIZING2014, maNoMissingPhotons2020, barrowLymanContinuumEscape2020, paardekooperFirstBillionYears2015,rosdahlLyCEscapeSPHINX2022,kostyukIonizingPhotonProduction2023, katzSphinxPublicData2023}. Nonetheless, to properly compare any distribution with our results, we need to identify simulations using: (i) a similar definition of the escape fraction as the one adopted in this work and (ii) a galaxy sample closely analogous to the final sample used for inferring the exponential distribution in this study.

Most simulations report the angle-averaged, cosmic escape fraction of LyC photons, while in this work, we use observational data and are thus limited to sight line-dependent escape fractions. The \texttt{SPHINX} suite of cosmological radiation-hydrodynamical simulations of reionisation \citep{katzSphinxPublicData2023}, however, provides the LyC escape fraction (measured at rest frame wavelength 900Å) for 10 different sight lines for each galaxy in the simulation and is therefore a suitable candidate for comparison with our results.

We make the following selection of \texttt{SPHINX} galaxies to best emulate our sample:
\begin{itemize}
    \item[$\bullet$] The \texttt{SPHINX} galaxies present redshifts spanning from $4.64<z<10$, while our sample exhibit redshifts in the range $3.35<z<3.95$. We select \texttt{SPHINX} galaxies with $z<5.5$ for the comparison.
    \item[$\bullet$] The \texttt{SPHINX} galaxies are generally lower mass than our sample displaying $6.51 < \log(M_*/M_\odot) < 10$, while our sample shows masses in the range $8.66 < \log(M_*/M_\odot) < 10.44$. We select all \texttt{SPHINX} galaxies within the mass range of our sample. To avoid an otherwise considerable overweight of lower mass galaxies compared to our sample, we further down-sample galaxies with $\log(M_*/M_\odot) < 9$ by 90 per cent (i.e. keeping 10 per cent).
    \item Each \texttt{SPHINX} galaxy presents 10 different sightlines. We select galaxies with sightlines where the observed optical UV $\beta$-slope is within the range of our galaxy sample such that $-2.68 < \beta_\mathrm{obs} < 0.43$. Note, however, that \texttt{SPHINX} galaxies are generally more bluer (exhibiting lower $\beta$-slopes) than our sample. 
\end{itemize}

\begin{figure}
    \centering
    \includegraphics[width=\textwidth/2]{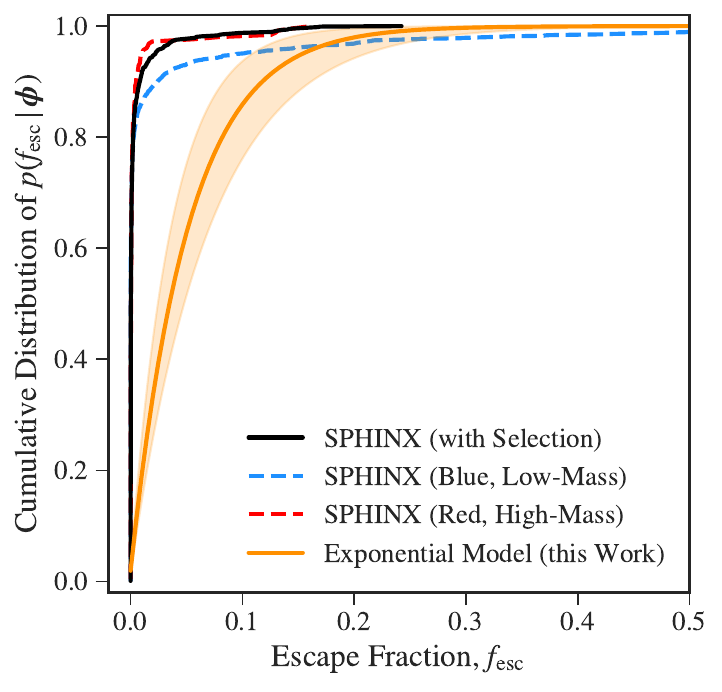}
    \caption{The cumulative distribution of the population distribution of the escape fraction. In orange, we display the median CDF for the exponential model inferred in this work (the shaded region marks the 16th and 84th percentile interval). The black line marks the empirical cumulative distribution function for a selection of \texttt{SPHINX} galaxies with somewhat similar properties to our observed sample. For context, we have plotted the CDF for selections of red, high-mass galaxies and blue, low-mass galaxies in \texttt{SPHINX}. All selections are described in Section \ref{sub:sim_comparison}).}
    \label{fig:sphinx_comparison}
\end{figure}

In Figure \ref{fig:sphinx_comparison}, we display the analytical CDF of the distribution of the LyC escape fraction as inferred from our exponential model along with the empirical CDF from 1139 \texttt{SPHINX} sightlines. While the shape of the CDFs hold some similarities, the distributions do not match and the simulations under-predict the escape fraction relative to our result.

Accounting for the differences in galaxy properties between the compared samples, the distributions are predicted to be less consistent than what is shown in Figure \ref{fig:sphinx_comparison}.
For one, the redshift range of our sample is lower than those exhibited by the \texttt{SPHINX} galaxies. \cite{rosdahlLyCEscapeSPHINX2022} predicts (using the \texttt{SPHINX} suite of simulations) a correlation between redshift and global escape fraction, indicating that simulated galaxies with lower redshifts would exhibit even lower escape fractions. In addition, the \texttt{SPHINX} galaxies exhibit an overweight of blue (lower UV $\beta$-slopes) compared to our sample, which suggests \texttt{SPHINX} galaxies should display a larger mean $\fesc$ compared to the mean of our inferred exponential distribution. 

In Figure \ref{fig:beta-split}, we demonstrated that bluer galaxies have an escape fraction distribution shifted to higher values. In one further attempt to match any selection of \texttt{SPHINX} galaxies to our inferred exponential distribution, we select blue, low-mass galaxies ($6.51 < \log(M_*/M_\odot) < 9.5$, $-2.68 < \beta_\mathrm{obs} < - 2$) and red, high-mass galaxies ($9.5 < \log(M_*/M_\odot) < 10.63$, $-2 < \beta_\mathrm{obs} < 0.68$) from the full \texttt{SPHINX} catalogue (still imposing $z<5.5$) and plot their empirical CDF (displayed in Figure \ref{sub:sim_comparison}). The \texttt{SPHINX} galaxies reproduce the trend found in this work, with blue galaxies exhibiting a distribution shifted towards larger escape fractions, but even the bluest galaxies in the \texttt{SPHINX} sample are not able to reproduce our inferred exponential distribution and still under-predicts the escape fraction.

To better compare simulations with observational constraints on the distribution of the escape fraction in the future, we encourage simulations to report sight-line, galactic-scale escape fractions for galaxies at lower redshift.
\section{Conclusions}
\label{sec:conclusion}
In this work, we use data from the VANDELS spectroscopic survey \citep{mclureVANDELSESOPublic2018, garilliVANDELSESOPublic2021} to measure the distribution of the ionising photon escape fraction from 148 star-forming, reionisation-era analogue galaxies ($3<z<4$). This sample of galaxies was originally presented in \cite{begleyVANDELSSurveyMeasurement2022}. 

We have developed a hierarchical Bayesian inference framework to infer the distribution of the ionising photon escape fraction from the ratio of LyC to non-ionising UV flux, as measured by broadband photometry. Our method incorporates careful treatment of the IGM and CGM transmission of ionising photons and accounts for the effects of dust attenuation in the UV and the uncertain intrinsic ratio of ionising to non-ionising UV flux. Within this framework, we have tested 3 physically motivated distributions beyond the constant model developed by \cite{begleyVANDELSSurveyMeasurement2022}: an exponential, a log-normal and a bimodal parameterisation of the population distribution of escape fractions.

The main results of this study are summarised as follows:
\begin{enumerate}[(i)]
    \item The constant, exponential and bimodal population distributions all recover the same expected value of $\fesc=5\%$, consistent with the result of \citet{begleyVANDELSSurveyMeasurement2022}, while the log-normal parameterisation finds an expected value of $29\%$.
    \item Using a posterior predictive test, we find that the exponential and constant distributions are best at reproducing the measured LyC to non-ionising UV flux ratios of our galaxy sample. The performance of the two distributions is, however, not statistically distinguishable (with the current sample size). The bimodal and log-normal distributions are ruled out with high significance.
    \item We argue that the exponential distribution is most likely representative of the true distribution of the escape fraction in our sample. Unlike the constant model, the exponential distribution is consistent with individual constraints on the escape fraction obtained for 2 galaxies in our sample, which place a lower limit on the escape fraction of $\fesc>0.38$ in these galaxies. 
    \item We find that the bimodal model of the distribution of the escape fraction in our sample is disfavoured, implying that a purely bimodal escape of ionising photons, via strong sightline and/or time variability, is improbable.
    \item We split our sample into two subsamples based on the UV Continuum slope ($\beta$-slope) and repeat the full hierarchical inference scheme to infer an exponential distribution representative of each sample. We find that the escape fraction anti-correlates with the $\beta$-slope, such that bluer galaxies (lower $\beta$-slopes) are more likely to populate the high-end tail of the escape fraction distribution. In particular, we infer the scale parameter for blue and red galaxies as, respectively, $\mu_\mathrm{blue}=\BlueExpMuVal$ and $\mu_\mathrm{red}=\RedExpMuVal$.
    \item We compare the inferred exponential distribution of the escape fraction (with scale parameter $\mu=\ExpMuVal$) to sight-line escape fractions reported from the \texttt{SPHINX} suite of cosmological radiation-hydrodynamical simulations of reionisation \cite{katzSphinxPublicData2023}, but we find that the simulations under-predict the escape fractions relative to the results from our sample. While the galaxies in \texttt{SPHINX} and our sample are not directly analogous, presenting (among other properties) different redshift ranges, based on the other predicted trends in the simulations even with a better-matched simulated sample the \texttt{SPHINX} distribution would remain inconsistent with our results.
\end{enumerate}

Reproducing the line-of-sight escape fraction distribution can be a key test for the implementation of physical conditions and feedback mechanisms in simulations. Future studies would benefit from
radiation-hydrodynamical simulations reporting line-of-sight specific escape fractions, in addition to the typically reported angle-averaged escape fractions. This would enable a more direct comparison with the observations and a better understanding of the physical processes responsible for the escape of ionising photons.

Our results also have important implications for the reionisation process itself. Models commonly assume a constant or deterministic escape fraction from galaxies during reionisation \citep[][]{madauRadiativeTransferClumpy1999,robertsonEarlyStarformingGalaxies2010,parkInferringAstrophysicsReionization2019,naiduRapidReionizationOligarchs2020,munozReionizationJWSTPhoton2024}. An exponential distribution implies that only a small fraction of sources may be contributing to reionisation at a given time, likely due to a combination of time and/or sightline variability. Thus, considering the escape fraction \textit{distribution} may alleviate the ionising photon `budget crisis' implied by recent JWST estimates of the ionising photon production efficiency 
\citep{munozReionizationJWSTPhoton2024}. 

We have also demonstrated that the shape of the $\fesc$ distribution depends on the UV $\beta$ slope, with bluer galaxies having a distribution shifted to higher $\fesc$.
With larger samples, it will be possible to constrain the escape fraction distribution parameters as a function of galaxy properties such as specific star formation rate, stellar mass, UV magnitude and $\beta$-slope. This dependence on galaxy properties will be important for predicting the escape fraction distribution during the Epoch of Reionisation.

\begin{acknowledgements}
      We thank Joki Rosdahl for useful discussions and sharing the SPHINX $\fesc$ catalog. KCK and CAM acknowledge support by the Carlsberg Foundation under grant CF22-1322. CAM also acknowledges support by the VILLUM FONDEN under grant 37459. The Cosmic Dawn Center (DAWN) is funded by the Danish National Research Foundation under grant DNRF140. 
      FC acknowledges support from a UKRI Frontier Research Guarantee Grant (PI Cullen; grant reference EP/X021025/1).\\
      This work made use of the following open-source software: \texttt{Matplotlib} \citep{hunterMatplotlib2DGraphics2007}, \texttt{Numpy} \citep{harrisArrayProgrammingNumPy2020}, \texttt{SciPy} \citep{virtanenSciPyFundamentalAlgorithms2020}, \texttt{Astropy} \citep{theastropycollaborationAstropyProjectSustaining2022}, \texttt{emcee} \cite{foreman-mackeyEmceeMCMCHammer2013}, \texttt{Pandas} \citep{thepandasdevelopmentteamPandasdevPandasPandas2024}, \texttt{Corner} \citep{foreman-mackeyCornerPyScatterplot2016}, and \texttt{Numba} \citep{lamNumbaLLVMbasedPython2015}.
\end{acknowledgements}

\bibliographystyle{aa} 
\bibliography{references.bib} 

\section*{Appendix A: Simulating Observed Flux Ratios}
\label{app:mocks}

When testing our inferred four distributions' ability to describe the observed data, i.e. the observed LyC to non-ionising UV flux ratios, $\{\Robs\}$, we use a simulation framework to produce a set of mock $\Robs$ values based on a given distribution of the escape fraction (See Section \ref{sub:results_comparison}). 

For each $\Robs$ value to produce (i.e. each mock galaxy), the simulation takes the following, fixed input parameters: a redshift ($z$), an observational error on $\Robs$ ($\sigma_\mathcal{R}$) and a value for dust attenuation in the V-band ($A_\mathrm{V}$). We sample these parameters from our catalogue of real galaxies used in this study. The galaxy sample and the measurement of the aforementioned parameters were originally presented in \cite{begleyVANDELSSurveyMeasurement2022}. 

In addition, the simulation requires an escape fraction ($\fesc$), an intrinsic flux ratio ($\Rint$) and an integrated transmission value ($\trans$). The escape fraction is sampled from the inferred distribution that is being tested. We first sample a set of population parameters ($\pop$) from the inferred posterior distribution, secondly, we compute the corresponding probability distribution of the escape fraction, and lastly, we sample an escape fraction from this distribution.

The intrinsic flux ratio, $\Rint$ is sampled from the corresponding prior distribution valid for the given redshift (See sections \ref{sub:rint} and \ref{sub:single_layer}). Similarly, the integrated transmission is sampled from its redshift-dependent prior (See sections \ref{sub:trans} and \ref{sub:single_layer})

To obtain the observed flux ratio of LyC to non-ionising UV flux, $\Robs$, of a simulated galaxy, we need to model its observed spectrum, from which we can directly compute $\Robs$. Recall that $\Robs$ is defined in terms of an average flux density per unit frequency (Equation \ref{eq:robs}). Armed with a spectrum, we compute the average flux density per unit frequency, $\langle f_\mathrm{LyC} \rangle$ and $\langle f_\mathrm{UV} \rangle$, by integrating the spectrum through the appropriate filter, at the redshift of the galaxy. The average flux density per unit frequency, $\langle f_\nu \rangle$, is defined as:
\begin{equation}
    \langle f_\nu \rangle = \frac{ \int \frac{f_{\nu}}{\nu} R_{\nu} d\nu  }{\int \frac{R_\nu}{\nu} d\nu}
    \label{eq:flux_density}
\end{equation}
where $R_\nu$ is the filter transmission function and $f_\nu$ is the flux density in frequency space. When we have estimated the average flux density for both the LyC filter and the UV filter, obtaining $\Robs$/$\Rint$ is just a matter of dividing the two.

The template for the intrinsic mock spectra is the Binary Population and Spectra Synthesis (BPASS) model \citep{eldridgeBinaryPopulationSpectral2017} that was fitted to the composite spectra of the real galaxy sample (see Section \ref{sub:rint} and \citealp{begleyVANDELSSurveyMeasurement2022}). The observed spectrum is obtained by redshifting the intrinsic spectrum and modifying the flux to account for the imposed escape fraction, dust attenuation and the optical depth of the IGM and CGM.

The full procedure for simulating the observed spectrum from the intrinsic BPASS model and measuring the corresponding $\Robs$ is detailed below:
\begin{enumerate}
    \item We convert the units from $L_\odot/\AA$ (units of the template spectrum, luminosity density in wavelength space) to $\mathbf{\mu}$Jy=erg/s/cm$\mathbf{^2}$/Hz (flux density in frequency space) and redshift the spectrum according to the input parameter, $z$.

    \item  Fluxes at wavelengths below the Lyman limit ($912$Å) are reduced by multiplying with the given escape fraction, $\fesc$.

    \item We apply dust attenuation at wavelengths above the Lyman limit using the \cite{reddySPECTROSCOPICMEASUREMENTSFARULTRAVIOLET2016} dust curve and the input value for the attenuation in the V-band, $A_\mathrm{V}$.

    \item We compute the flux ratio using the definition of $\Robs$ in Equation \ref{eq:robs}, where the average flux density per unit frequency is obtained with Equation \ref{eq:flux_density}.

    \item We multiply the flux ratio from step (4) according to the integrated LyC transmission ($\trans(z)$). 

    \item To account for the intrinsic flux ratio, we multiply the ratio from step (5) with the sampled input $\Rint$ divided by the intrinsic flux ratio of the original BPASS spectrum.

    \item Finally, we sample an error on $\Robs$ from a normal distribution with standard deviation given by the input $\sigma_R$ and add it to the reported $\Robs$ value.
\end{enumerate}

\end{document}